# Effect of the Earth's Time-Retarded Transverse Gravitational Field on Spacecraft Flybys
### (Version 3)

## J. C. Hafele[1]

In 2008 a group of NASA scientists reported an anomalous change in the speed of six spacecraft flybys of the Earth. The reported speed change for the NEAR spacecraft flyby is 13.46±0.01 mm/s. It is known that general relativity theory reduces to classical time-retarded electromagnetic field theory in the linearized approximation. This report shows that time-retarded field theory applied to the Earth's transverse gravitational field gives rise to a small change in the speed of a spacecraft during a flyby. The speed change depends on the time dependence of the transverse gravitational field, which generates an induction field that is proportional to the time derivative of the transverse gravitational field. If the corresponding speed for the induction field is set equal to (4.130±0.003) times the Earth's equatorial rotational surface speed, and the speed of the Earth's gravitational field is set equal to (1.060±0.001) times the vacuum speed of light, the calculated value for the speed change for the NEAR flyby agrees exactly with the observed value. Similar results are found for the remaining five flybys.

## 1. INTRODUCTION

It has long been known that electromagnetic fields propagate at the speed of light. The actual speed of light depends on whether the field is propagating in a vacuum or in a material medium. In either case, to calculate the electromagnetic fields of a moving point-charge, the concept of "time retardation" must be used.[1] Time retardation is necessary because it takes a certain amount of time for causal physical fields to propagate from a moving point-source to a distant field point.

Unlike the speed of light, the "speed of gravity" in general relativity theory is difficult to comprehend and even more difficult to measure. There is ample evidence for the belief that the speed at which gravitational radiation propagates in empty space is the same as the vacuum speed of light.[2] But the speed at which the gravitational field of a massive central object propagates may differ slightly from the vacuum speed of light. In 1898, more than 18 years before Einstein developed the final version of his general relativity theory, the speed of the Sun's gravity field was found by a high school math teacher, Paul Gerber, by calculating what it would need to be to cause the anomalous advance of the perihelion of Mercury.[3] Gerber's value, $3.05500 \times 10^8$ m/s, is about 2% greater than the vacuum speed of light.

---

1. 618 S. 24th St., Laramie, WY 82070, USA, e-mail: cahafele@bresnan.net

In 2002 a group of radio astronomers measured the speed of Jupiter's "gravitational field" by detecting the rate of gravitational bending of radio waves from a distant quasar as the giant planet crossed the line-of-sight.[4] They concluded that the speed of Jupiter's "gravitational field" is 1.06±0.21 times the vacuum speed of light. Though controversial, these results suggest that the speed of propagation of the gravitational field of a massive central object may not be exactly the same as the vacuum speed of light.

The first terrestrial measurement that proved a connection between gravity and light, the gravitational red-shift, was carried out by Pound and Rebka in 1959.[5] In 1972 Hafele[2] and Keating reported the results of their experiments which detected the relativistic time dilation and the gravitational red-shift for precision clocks flown around the world using commercial jet flights.[6] These experiments show conclusively that clocks at a deeper gravitational potential run slower and that moving clocks run slower. To correct for these relativistic effects, the precision clocks used in the GPS system are adjusted before they are launched into space.

In 2008 a group of NASA scientists reported a small anomalous change in the orbital energy of spacecrafts after flybys of the Earth.[7,8] The observed anomalous change in speed for six flybys range between +14 and -5 mm/s. An empirical formula developed by J. D. Anderson et al.[7] gives a value of 13.28 mm/s for the NEAR spacecraft flyby, which is about 1.2% less than the observed value of 13.46±0.01 mm/s. The empirical formula predicts that the speed change is proportional to the ratio of the Earth's equatorial rotational speed to the vacuum speed of light, and depends on the difference between the inbound and outbound declinations.

In 2009 R. T. Cahill reported a new "dynamical theory of 3-space" that explains numerous light-speed anisotropy experiments.[9] Cahill used his new theory to predict a value of 13.45 mm/s for the NEAR flyby, which agrees very well with the observed value of 13.46±0.01 mm/s. Agreement with the other flybys is not this good. Cahill's theory produces results that depend on a universal anisotropy in the speed of light.

The objective of this report is to provide an alternative theory that explains flyby anomalies by showing that the speed change may be caused by time retardation in the Earth's transverse gravitational field. This neoclassical time-retarded theory produces results that depend on the time-dependence of the transverse gravitational field, which in turn depends on the ratio of the Earth's equatorial rotational speed to the speed of the Earth's gravitational field.

The concepts of "field" and "time retardation" do not exist in general relativity theory, but they do emerge in the linearized version of general relativity theory. In 2002 F. Rohrlich derived a time-retarded version for Newton's law of gravity, a version that satisfies causality, by using a first-order approximation for the space-time curvature.[10] Rohrlich based his analysis on the linear approximation to general relativity theory in the popular textbook by W. Rindler.[11] Rindler shows in great detail that general relativity theory reduces in the linear approximation to classical time-retarded electromagnetic field theory. The resulting formulas for the time-retarded Newtonian gravitational field of a moving point-mass are used herein to derive a first-order approximation for the effect of time retardation on the gravitational field of a large spinning sphere during spacecraft flybys.

________________

2. The author of this report.

Induction fields are well-known in classical electromagnetic field theory.[1] The neoclassical time-retarded theory proposed herein requires a fundamental hypothesis: the time-dependence of the transverse gravitational field generates an induction field that is proportional to the time derivative of the transverse field.

In Sec. 2, the linear approximation for general relativity theory is reviewed, the time-retarded transverse gravitational field for a circulating point mass is derived, the empirical prediction formula found by J. D. Anderson et al.[7] is reviewed, and the procedure for calculating the time-retarded transverse gravitational field of a large rotating sphere is developed. The formula for the time-retarded transverse gravitational field requires a triple integration over the volume of the central sphere. Solution of the triple integral can be accomplished by numerical integration, but it takes a lot of computer time. To speed up the calculation, an equivalent power series is substituted for the triple integration. To calculate the induction field requires a fourth integration, and to calculate the speed change requires a fifth integration. These final integrals are solved by using numerical integration.

In Sec. 3, the reported trajectory parameters for the NEAR spacecraft are used to calculate the time dependence of the spacecraft's radial distance, velocity, and angular speed. Then the formulas needed to calculate the time-retarded transverse gravitational field are developed.

In Sec. 4, the central sphere is divided into a sequence of infinitesimal volume elements and corresponding elemental point masses. Next the retarded distance between a circulating point mass inside the rotating sphere and a field point outside the sphere, and the retarded-time derivative of the retarded distance, are found. Then the formulas for the transverse field, the time derivative of the transverse field, and the induction field for the NEAR flyby are found. If the induction speed is set equal to $4.130\pm0.003$ times the Earth's equatorial surface speed and the speed of the Earth's gravity field is set equal to $1.060\pm0.001$ times the vacuum speed of light, the calculated value for the anomalous change in speed for the NEAR flyby is $13.46\pm0.01$ mm/s, in exact agreement with the observed value.

In Sec. 5, a comparison of the observed speed change for all six flybys with the corresponding calculated speed change from the time-retarded theory, from a modified empirical formula, and from Cahill's anisotropic light-speed theory is presented. A useful correlation between the speed change predicted by the time-retarded theory and the eccentricity for the flyby trajectory is found. Any preference for the various theories is left to the reader.

## 2. TIME-RETARDED TRANSVERSE GRAVITATIONAL FIELD AND INDUCTION FIELD OF A CENTRAL ROTATING OBJECT

### 2.1 Linear Approximation for General Relativity Theory

The linear approximation for general relativity theory provides a valid first-order approximation for the gravitational field of a moving point mass. The linear approximation applies for "slowly" moving particles in "weak" gravitational fields. The word "slowly" means $|u| \ll c$, where $|u|$ is the maximum absolute magnitude for the particle speed, and the word "weak" means $|\varphi| \ll c^2$, where $|\varphi|$ is the maximum absolute magnitude for the gravitational potential.

The section entitled **The Linear Approximation to GR** in the popular textbook by W. Rindler[11] starts on page 188. The following is a direct quote from pages 190 and 191.

In the general case, Equations (**8.180**) can be integrated by standard methods. For example, the first yields as the physically relevant solution,

$$\gamma_{\mu\nu} = -\frac{4G}{c^4} \iiint \frac{\left[ T_{\mu\nu} \right] dV}{r} \quad , \tag{8.184}$$

where [] denotes the value "retarded" by the light travel time to the origin of r.

As an example, consider a system of sources in stationary motion (e.g., a rotating mass shell). All $\gamma$'s will then be time-independent. If we neglect stresses and products of source velocities (which is not really quite legitimate[14]), the energy tensor (**8.128**) becomes

$$T_{\mu\nu} = \begin{pmatrix} \mathbf{0}_3 & -c^2 \mathbf{v} \\ -c^2 \mathbf{v} & c^4 \rho \end{pmatrix} \quad , \tag{8.185}$$

where $\mathbf{0}_3$ stands for the 3×3 zero matrix, and so, from (**8.184**),

$$\gamma_{ij} = 0, \qquad (i, j = 1, 2, 3) \quad . \tag{8.186}$$

For slowly moving test particles, ds=cdt. If we denote differentiation with respect to t by dots, the first three geodesic equations of motion become [cf. (**8.15**)]

$$\ddot{x}^i = -\Gamma^i_{\mu\nu} \dot{x}^\mu \dot{x}^\nu \tag{8.187}$$

$$= -\left(\gamma^i{}_{\mu,\nu} - \tfrac{1}{2}\gamma_{\mu\nu,}{}^i - \tfrac{1}{4}\eta^i{}_\mu\gamma_{,\nu} - \tfrac{1}{4}\eta^i{}_\nu\gamma_{,\mu} + \tfrac{1}{4}\eta_{\mu\nu}\gamma_{,}{}^i\right)\dot{x}^\mu\dot{x}^\nu \quad , \tag{8.188}$$

where we have substituted into (**8.187**) from (**8.168**) and (**8.172**) and used $\gamma = \eta^{\mu\nu}\gamma_{\mu\nu} = -h$. Moreover, $\gamma = c^{-2}\gamma_{44}$. Now if we let $\dot{x}^\mu = (u^i, 1)$ and neglect products of the u's, Equation (**8.188**) reduces to

$$\ddot{x}^i = -\gamma^i{}_{4,j}u^j + \gamma_{j4,}{}^i u^j + \tfrac{1}{4}\gamma_{44,}{}^i \quad .$$

This can be written vectorially in the form

$$\ddot{\mathbf{r}} = \mathbf{grad}\varphi - \frac{1}{c}\left(\mathbf{u} \times \mathbf{curl\ a}\right) = -\left[\mathbf{e} + \frac{1}{c}\left(\mathbf{u} \times \mathbf{h}\right)\right] \tag{8.189}$$

where [cf.(**8.184**),(**8.185**)]

$$\varphi = -\frac{1}{4}\gamma_{44} = G\iiint \frac{[\rho]\,dV}{r} \quad , \qquad \mathbf{a} = -\frac{c}{4}\gamma^i{}_4 = \frac{1}{c}G\iiint\frac{[\rho\mathbf{u}]\,dV}{r} \quad , \tag{8.190}$$

and

$$\mathbf{e} = -\mathbf{grad}\varphi \quad , \qquad \mathbf{h} = \mathbf{curl}\ 4\mathbf{a} \quad . \tag{8.191}$$

The formal similarity with Maxwell's theory is striking. The only differences are: the minus sign in (**8.189**)(because the force is attractive); the factor G in (**8.190**)(due to the choice of units); and the novel factor 4 in (**8.191**)(ii).

The formula for $\varphi$, Eq. (**8.190**), gives the time-retarded Newtonian gravitational potential, and the formula for **e**, Eq. (**8.191**), gives the time-retarded Newtonian gravitational field in the linear approximation to general relativity theory.

The following is a direct quote from page 413 of Rohrlich's paper[10] entitled: **Causality, the Coulomb field, and Newton's law of gravitation.**

Historians tell us that Newton was quite unhappy over the fact that his law of gravitation implies an action-at-a-distance interaction over very large distances such as that between the sun and the earth. But he was unable to resolve this problem.[6] With the aid of general relativity, one can show that even a first-order correction to his law produces a causal interaction. I shall only sketch how this result arises, and I refer the reader to the excellent text by Rindler[7] or to other texts on general relativity for further details.

Newton's law of gravitation,

$$F = GMm/r^2 \tag{4.1}$$

violates causality, because it does not take into account that it takes a certain time for the interaction to travel from the source M to the mass m on which it acts. In Einstein's theory of gravitation (general relativity), interactions propagate with the speed of light, and gravitation is the result of a curvature of space-time due to the source M rather than the force. Its field equations are

$$R_{\mu\nu} = \kappa \left( T_{\mu\nu} - \tfrac{1}{2} g_{\mu\nu} T \right) \quad . \tag{4.2}$$

Here, $R_{\mu\nu}$ is the Ricci tensor (a contraction of the curvature tensor), $\kappa$ is the coupling constant, $\kappa = 8\pi G$, and the matter tensor is assumed to be that of dust, $T_{\mu\nu}(x) = \rho u_{\mu}(x) u_{\nu}(x)$.

Because curvature depends on the derivatives of the metric tensor, $g_{\mu\nu}$, Einstein gravitation can be described to first order as small deviations from the metric for flat space-time, $\eta_{\mu\nu}$, so that

$$g_{\mu\nu} = \eta_{\mu\nu} + h_{\mu\nu} \quad , \tag{4.3}$$

where $h_{\mu\nu}$ is small compared to $\eta_{\mu\nu}$. In this approximation, the field equation (4.2) becomes[7]

$$\Box h_{\mu\nu} = -2\kappa \left( T_{\mu\nu} - \tfrac{1}{2} \eta_{\mu\nu} T \right) \quad . \tag{4.4}$$

As in the electromagnetic case, one can impose causality by choosing the retarded solution,

$$h_{\mu\nu}(x) = 2\kappa \int D_R(x - x') \left[ T_{\mu\nu}(x') - \tfrac{1}{2} \eta_{\mu\nu}(x') T(x') \right] d^4 x' \quad . \tag{4.5}$$

To show that this equation reproduces the equation of Newtonian gravitation theory, consider its static limit. In this limit, all components of $T_{\mu\nu}$ vanish except $T_{00} = \rho$. Equation (4.4) then reduces to (again using c=1),

$$\nabla^2 h_{00} = -8\pi G\rho \quad , \tag{4.6}$$

while all the other $h_{\mu\nu}$ vanish. Equation (4.6) is of course exactly the equation for the Newtonian gravitational potential $\varphi$. But because the Newtonian and Einsteinian gravitation theories have completely different pictures of gravitation, the symbols in the equation must be reinterpreted.[8] Thus, the component $h_{00}$ of the correction to the Minkowski metric tensor is to be interpreted as the Newtonian potential (except for a factor of 2), $h_{00} = 2\varphi$. Equation (4.6) then becomes

$$\nabla^2 \varphi = -4\pi G\rho \quad , \tag{4.7}$$

and the Newtonian gravitational force acting on a mass m is $\mathbf{F} = m\boldsymbol{\nabla}\varphi$.[9] (Note the gravitational potential has different dimensions than the electric one.)

Of course, one could play the same trick (3.14) as in the electromagnetic case. But it seems without justification here. Instead, let us return to Eq. (4.5), the *solution* of the differential equation for $h_{\mu\nu}$. We see that the *solution is retarded*. If we take the Newtonian limit, we find the desired result,

$$h_{00} = -16\pi G \int D_R(x - x') [\rho(x')/2] d^4x' \ .$$

The Newtonian gravitational potential is therefore

$$\varphi = h_{00}/2 = -4\pi G \int \rho(t - |\mathbf{x} - \mathbf{x}'|, \mathbf{x}') d^3x'/|\mathbf{x} - \mathbf{x}'| \ , \qquad (4.8)$$

so that the Newtonian gravitational force $\mathbf{F} = m\nabla\varphi$ is now *retarded*. By keeping the time derivatives in the differential equation for $h_{00}$, the requirement for causality can be met. Because the Newtonian theory is entirely static, retardation is not possible until the correction due to deviations from Minkowski space is considered.

Thus we find that both Rindler and Rohrlich arrive at the same conclusion. In the limit of slow speeds and weak fields, i.e., the linearized approximation, general relativity theory reduces to classical <u>time-retarded</u> electromagnetic <u>field</u> theory, and the gravitational field generated by a moving point mass can be found to the first-order of approximation by retarding the position of the source by the time it takes for the interaction to propagate from the source to the field point.

The gravitational fields that are analogous to magnetic fields, "gravitomagnetic fields", result from the vector $\mathbf{a}$ in Rindler's formula Eq. (**8.190**), where the speed c appears explicitly. Substituting the formula for $\mathbf{a}$, Eq. (**8.190**), into the formula for $\ddot{\mathbf{r}}$, Eq. (**8.189**), gives

$$\ddot{\mathbf{r}} = \mathbf{grad}\varphi - \frac{1}{c}\left(\mathbf{u} \times \mathbf{curl}\left(\frac{1}{c} G\iiint \frac{[\rho\mathbf{u}]\,dV}{r}\right)\right) \ .$$

By substituting $\mathbf{u}/u$ for $\mathbf{u}$, the formula for $\ddot{\mathbf{r}}$ becomes

$$\ddot{\mathbf{r}} = \mathbf{grad}\varphi - \frac{u^2}{c^2}\left(\frac{\mathbf{u}}{u} \times \mathbf{curl}\left(G\iiint \frac{\left[\rho\frac{\mathbf{u}}{u}\right]dV}{r}\right)\right) \ .$$

This shows that fields resulting from the curl of $\rho\mathbf{u}/r$ are of order $u^2/c^2$ and higher. These fields will not be studied herein because the objective is to seek the time-retarded gravitational field of a large rotating sphere to the first order in $u/c$.

To avoid confusion between $\varphi$ and the azimuthal angle $\phi$, the symbol for gravitational scalar potential will be changed to $\chi$. The traditional symbol for gravitational acceleration fields is $\mathbf{g}$ (boldface usually denotes a vector). Notice that $\chi$ is a scalar function and $\mathbf{g}$ is a vector function. (The reader is cautioned not to confuse the vector gravitational field, $\mathbf{g}$, with the elements of the metric tensor, $g_{\mu\nu}$.) The time-retarded Newtonian gravitational scalar potential $\chi$ and the time-retarded Newtonian gravitational field **g** for a circulating point-mass of mass m can now be written to first order in u/c,

$$\chi = -\frac{Gm}{r'} \quad , \qquad\qquad \mathbf{g} = -\mathbf{grad}\chi = -\frac{Gm}{r'^2}\frac{\mathbf{r'}}{r'} \quad , \qquad (2.1.1)$$

where r' is the magnitude for the time-retarded distance from the moving point mass to the field point, and **r'**/r' is a unit vector directed towards increasing r'. Notice that the speed c does not appear explicitly in Eq. (2.1.1), but it is implicit in the time-retarded distance r'.

This <u>neoclassical time-retarded Newtonian gravitational field theory</u> is based on a solid theoretical foundation that results from the linear approximation for general relativity theory. It goes one small step beyond the classical Newtonian <u>instantaneous</u> action-at-a-distance theory, and represents the ultimate first-order approximation for slow speeds and weak fields. The instantaneous classical Newtonian theory has been used successfully for more than three hundred years, so it probably is not totally wrong. This neoclassical time-retarded version produces a small correction and is good only as a first-order approximation. It cannot be expected to reproduce effects that are of order $u^2/c^2$ and higher, such as time dilation and space contraction, the relativistic advance of the perihelia of the planets, the relativistic gravitational bending of starlight, the Lense-Thirring effect, gravitational radiation from colliding black holes, worm holes between parallel universes, etc.

## 2.2 Time-Retarded Transverse Field of a Circulating Point Mass

Let's do a simple thought experiment using classical Newtonian concepts. Consider a point mass of very small mass m that is constrained by nongravitational forces to circulate around the origin at a fixed radius R with a constant angular speed $\Omega'$, as depicted in Fig. 1. Let **g** be the gravitational field of m at a field point in the (x,y) plane which lies at the altitude h outside the circle. (The reader is cautioned not to confuse the altitude h with the elements of the deviation metric tensor, $h_{\mu\nu}$.) The space-time coordinates for the point-mass source are $(R,\phi',t')$. The field-point may or may not be constrained by a nongravitational force. The coordinates for the field-point are $(R+h,\phi,t)$. Let r' be the retarded distance between m and the field point. Let $g_h$ be the radial component and let $g_e$ be the transverse component of **g**. (The subscript "e" suggests that this component, if negative, is directed towards the east.)

A gravitational signal that is emitted by m at the retarded time t' will arrive at the field point at a slightly later time t. If the signal propagates at the speed $c_g$, $t=t'+r'/c_g$. If the point mass is moving very slowly, the signed magnitude of **g** is given by the neoclassical time-retarded version of Newton's law (Eq. (2.1.1)), $-Gm/r'^2$.

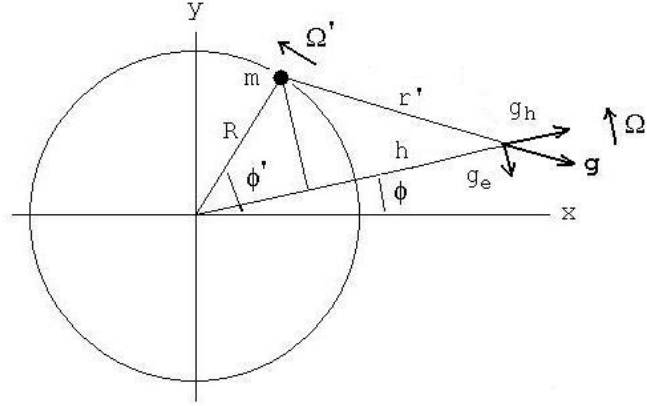

Figure 1. Radial and transverse components, $g_h$ and $g_e$, for the gravitational field **g** generated by a point mass m circulating around the origin at a fixed radius R with constant angular speed $\Omega'$.

The time dependence of the components $g_h$ and $g_e$ depends on the time dependence of the radial and transverse components of r'. Needed formulas are as follows.

$$t = t' + \frac{r'}{c_g} \quad , \qquad\qquad \frac{dt}{dt'} = 1 + \frac{1}{c_g}\frac{dr'}{dt'} \quad , \tag{2.2.1}$$

$$\phi' = \Omega't' \quad , \qquad\qquad \frac{d\phi'}{dt'} = \Omega' \quad , \tag{2.2.2}$$

$$\phi = \Omega t \quad , \qquad\qquad \frac{d\phi}{dt} = \Omega \quad , \tag{2.2.3}$$

$$\frac{dh}{dt} = v_h , \qquad\qquad (R + h)\frac{d\phi}{dt} = v_\phi , \tag{2.2.4}$$

$$r'_x = (R + h)\cos(\phi) - R\cos(\phi') \quad , \tag{2.2.5}$$

$$r'_y = R\sin(\phi') - (R + h)\sin(\phi) \quad , \tag{2.2.6}$$

$$r'_z = 0 \quad . \tag{2.2.7}$$

The magnitude for the retarded distance r' is

$$r' = \left(r'^2_x + r'^2_y\right)^{\frac{1}{2}} = \left((R + h)^2 + R^2 - 2R(R + h)(\cos(\phi)\cos(\phi') + \sin(\phi)\sin(\phi'))\right)^{\frac{1}{2}} \quad .$$

By using the trig identity for $\cos(\phi{-}\phi')$, the formula for r' reduces to

$$r' = \left((R + h)^2 + R^2 - 2R(R + h)\cos(\phi - \phi')\right)^{\frac{1}{2}} \quad . \tag{2.2.8}$$

The formula for the retarded-time derivative of r', to first order in $d\phi/dt$, is

$$\frac{dr'}{dt'} = \frac{1}{r'}R(R + h)\left(\frac{d\phi}{dt} - \frac{d\phi'}{dt'}\right)\sin(\phi - \phi') = \frac{R + h}{r'}R(\Omega - \Omega')\sin(\phi - \phi') \quad . \tag{2.2.9}$$

The radial and transverse components of **g** can be read directly from Fig. 1. If (x',y') is the frame which rotates with angular speed $\Omega$, the component $g_h$ lies parallel to the x'-axis and the component $g_e$ lies parallel to the y'-axis. The relative radial component and the relative transverse component of r' are

$$\text{radial component} \equiv RC = \frac{r'_x}{r'} = \frac{R + h - R\cos(\phi - \phi')}{r'} \quad , \tag{2.2.10}$$

$$\text{transverse component} \equiv TC = \frac{r'_y}{r'} = \frac{R\sin(\phi - \phi')}{r'} \quad . \tag{2.2.11}$$

There is a generalized method for finding the radial and transverse components. Let **R** be the radial vector to the source point, let **r** be the radial vector to the field point, and let **r'**≡**r-R**. Then the radial component is given by the vector dot product, **r∙r'**/rr', and the transverse component is given by the z-axis component of the vector cross product, $(\mathbf{r} \times \mathbf{r'})_z$/rr'. (The formula for the cross product can be found in most math handbooks.)

If the magnitude of a vector is known, and the direction angles relative to a system of orthogonal axes are known, then the components of the vector are known.

$$g_h = (g)\big(\text{radial component}\big) = (g)\big(RC\big)$$

$$= -\frac{Gm}{r'^2}\frac{R + h - R\cos(\phi - \phi')}{r'} = -\frac{Gm}{R^2}\frac{R^2\big(R + h - R\cos(\phi - \phi')\big)}{r'^3} \quad , \tag{2.2.12}$$

$$g_e = (g)\big(\text{transverse component}\big) = (g)\big(TC\big)$$

$$= -\frac{Gm}{r'^2}\frac{R\sin(\phi - \phi')}{r'} = -\frac{Gm}{R^2}\frac{R^3\sin(\phi - \phi')}{r'^3} \quad . \tag{2.2.13}$$

Now we can calculate time-average values. Let T' be the period for the circulating point mass, $T'=2\pi/\Omega'$. To simplify this thought experiment, assume for the moment that the field point is at rest in the nonrotating frame, that is, $v_h=0$, $\Omega=0$ and $\phi=0$. In this case, the formulas for r', dr'/dt', $g_h$, and $g_e$ reduce to the following.

$$r' = \Big((R + h)^2 + R^2 - 2(R + h)R\cos(\Omega't')\Big)^{\frac{1}{2}} \quad , \tag{2.2.14}$$

$$\frac{dr'}{dt'} = \frac{R + h}{r'}R\Omega'\sin(\Omega't') \quad , \tag{2.2.15}$$

$$g_h = -\frac{Gm}{R^2}\frac{R^2\big(R + h - R\cos(\Omega't')\big)}{r'^3} \quad , \tag{2.2.16}$$

$$g_e = -\frac{Gm}{R^2}\frac{R^3\sin(\Omega't')}{r'^3} \quad . \tag{2.2.17}$$

It is easy to show that a small part of $g_e$ is asymmetrical over the period T'. Let $\overline{g}_e^+$ be the time-average value for the positive half of the period, and let $\overline{g}_e^-$ be the time-average value for the negative half of the period, defined as follows.

$$\overline{g}_e^+ \equiv \int\limits_0^{T'/2} g_e(t') \frac{dt}{T'} = -\frac{Gm}{R^2} \int\limits_0^{T'/2} \frac{R^3 \sin(\Omega't')}{r'^3} \frac{dt}{T'} \ ,$$

$$\overline{g}_e^- \equiv \int\limits_{-T'/2}^0 g_e(t') \frac{dt}{T'} = -\frac{Gm}{R^2} \int\limits_{-T'/2}^0 \frac{R^3 \sin(\Omega't')}{r'^3} \frac{dt}{T'} \ . \tag{2.2.18}$$

Notice that the integration variable is t, but the integrand is a function of t'. This incompatibility can be removed by changing the integration variable from t to t'. Equation (2.2.1) gives the functional relationship t(t')=t'+r'/c$_g$. The transformation from t to t' can be accomplished by using the chain rule for differentiation,

$$\overline{g}_e^+ = -\frac{Gm}{R^2} \int\limits_{t'(0)}^{t'(T'/2)} \frac{R^3 \sin(\Omega't')}{r'^3} \frac{dt}{dt'} \frac{dt'}{T'} \ ,$$

$$\overline{g}_e^- = -\frac{Gm}{R^2} \int\limits_{t'(-T'/2)}^{t'(0)} \frac{R^3 \sin(\Omega't')}{r'^3} \frac{dt}{dt'} \frac{dt'}{T'} \ . \tag{2.2.19}$$

The derivative dt/dt' is by definition the Jacobian for the transformation from t to t'. In this case, the Jacobian is given by Eq. (2.2.15),

$$\text{Jacobian} = \frac{dt}{dt'} = 1 + \frac{1}{c_g} \frac{dr'}{dt'} = 1 + \frac{R\Omega'}{c_g} \frac{(R+h)}{r'} \sin(\Omega't') \ . \tag{2.2.20}$$

Notice that the effects of time retardation are found by using the Jacobian-1,

$$\text{Jacobian} - 1 = \frac{1}{c_g} \frac{dr'}{dt'} = \frac{R\Omega'}{c_g} \frac{(R+h)}{r'} \sin(\Omega't') \ . \tag{2.2.21}$$

Substituting the Jacobian into Eq. (2.2.19) gives the following formulas.

$$\overline{g}_e^+ = -\frac{Gm}{R^2} \int\limits_{t'(0)}^{t'(T'/2)} \frac{R^3}{r'^3} \sin(\Omega't') \frac{dt'}{T'} - \frac{Gm}{R^2} \frac{R\Omega'}{c_g} \int\limits_{t'(0)}^{t'(T'/2)} \frac{R^3(R+h)}{r'^4} \sin^2(\Omega't') \frac{dt'}{T'} \ ,$$

$$\overline{g}_e^- = -\frac{Gm}{R^2} \int\limits_{t'(-T'/2)}^{t'(0)} \frac{R^3}{r'^3} \sin(\Omega't') \frac{dt'}{T'} - \frac{Gm}{R^2} \frac{R\Omega'}{c_g} \int\limits_{t'(-T'/2)}^{t'(0)} \frac{R^3(R+h)}{r'^4} \sin^2(\Omega't') \frac{dt'}{T'} \ .$$

$$\tag{2.2.22}$$

Even functions are defined: f(x)=f(-x); odd functions: f(x)=-f(-x). Two examples for even functions are: cos(x)=cos(-x), $\sin^2(x)=\sin^2(-x)$. An example for an odd function: sin(x)=-sin(-x).

Let $\bar{g}_e$ be the total average value for $g_e$. The total is given by the sum of the two equations (Eq. (2.2.22)).

$$\bar{g}_e = \bar{g}_e^+ + \bar{g}_e^-$$

$$= -\left(\begin{array}{l} \dfrac{Gm}{R^2} \displaystyle\int_{t'(0)}^{t'(T'/2)} \dfrac{R^3}{r'^3}\sin\left(\Omega't'\right)\dfrac{dt'}{T'} \\[4mm] + \dfrac{Gm}{R^2} \displaystyle\int_{t'(-T'/2)}^{t'(0)} \dfrac{R^3}{r'^3}\sin\left(\Omega't'\right)\dfrac{dt'}{T'} \end{array}\right) - \left(\begin{array}{l} \dfrac{Gm}{R^2}\dfrac{R\Omega'}{c_g} \displaystyle\int_{t'(0)}^{t'(T'/2)} \dfrac{R^3\left(R+h\right)}{r'^4}\sin^2(\Omega't')\dfrac{dt'}{T'} \\[4mm] + \dfrac{Gm}{R^2}\dfrac{R\Omega'}{c_g} \displaystyle\int_{t'(-T'/2)}^{t'(0)} \dfrac{R^3\left(R+h\right)}{r'^4}\sin^2(\Omega't')\dfrac{dt'}{T'} \end{array}\right) .$$

$$(2.2.23)$$

Because the integrand for the first term (in parentheses) is an odd function, this term will cancel out (vanish), but because the integrand for the second or time-retarded term is an even function, the second term will not cancel out. So the formula for the residual time-average value for $\bar{g}_e$ reduces to

$$\bar{g}_e = -\frac{Gm}{R^2}\frac{R\Omega'}{c_g}\int_{t'(-T'/2)}^{t'(T'/2)}\frac{R^3\left(R+h\right)\sin^2(\Omega't')}{r'^4}\frac{dt'}{T'}$$

$$= -\frac{Gm}{R^2}\frac{R\Omega'}{c_g}\int_{-T'/2}^{T'/2}\frac{R^3\left(R+h\right)\sin^2(\Omega't')}{r'^4}\frac{dt'}{T'} \;, \qquad (2.2.24)$$

where the limits have been changed from $(t'(-T'/2)\to t'(T'/2))$ to $(-T'/2\to T'/2)$ for later convenience.

Because of the factor $R\Omega'/c_g$, the residual value for $\bar{g}_e$ will be very small, but however small it is, it will depend on the speed ratio $R\Omega'/c_g$.

Consider the average value for the radial component $g_h$. Let $\bar{g}_h^+$ and $\bar{g}_h^-$ be the time-average values for the radial component.

$$\bar{g}_h^+ = -\left(\frac{Gm}{R^2}\int_{t'(0)}^{t'(T'/2)}\frac{R^2}{r'^3}\left(R+h-R\cos\left(\Omega't'\right)\right)\frac{dt'}{T'}\right)$$

$$- \left(\frac{Gm}{R^2}\frac{R\Omega'}{c_g}\int_{t'(0)}^{t'(T'/2)}\frac{R^2\left(R+h-R\cos\left(\Omega't'\right)\right)\left(R+h\right)}{r'^4}\sin\left(\Omega't'\right)\frac{dt'}{T'}\right) \;,$$

$$\bar{g}_h^- = -\left(\frac{Gm}{R^2}\int_{t'(-T'/2)}^{t'(0)}\frac{R^2}{r'^3}\left(R+h-R\cos\left(\Omega't'\right)\right)\frac{dt'}{T'}\right)$$

$$- \left(\frac{Gm}{R^2}\frac{R\Omega'}{c_g}\int_{t'(-T'/2)}^{t'(0)}\frac{R^2\left(R+h-R\cos\left(\Omega't'\right)\right)\left(R+h\right)}{r'^4}\sin\left(\Omega't'\right)\frac{dt'}{T'}\right) \;.$$

Let $\bar{g}_h$ be the sum,

$$\bar{g}_h = \bar{g}_h^+ + \bar{g}_h^-$$

$$
\begin{aligned}
= &-\left( \frac{Gm}{R^2} \int_{t'(0)}^{t'(T'/2)} \frac{R^2}{r'^3} \left( R + h - R\cos\left(\Omega't'\right)\right) \frac{dt'}{T'} \right. \\
&\left. + \frac{Gm}{R^2} \int_{t'(-T'/2)}^{t'(0)} \frac{R^2}{r'^3} \left( R + h - R\cos\left(\Omega't'\right)\right) \frac{dt'}{T'} \right) \\
&-\left( \frac{Gm}{R^2} \frac{R\Omega'}{c_g} \int_{t'(0)}^{t'(T'/2)} \frac{R^2 \left( R + h - R\cos\left(\Omega't'\right)\right)\left( R + h\right)}{r'^4} \sin\left(\Omega't'\right) \frac{dt'}{T'} \right. \\
&\left. + \frac{Gm}{R^2} \frac{R\Omega'}{c_g} \int_{t'(-T'/2)}^{t'(0)} \frac{R^2 \left( R + h - R\cos\left(\Omega't'\right)\right)\left( R + h\right)}{r'^4} \sin\left(\Omega't'\right) \frac{dt'}{T'} \right) .
\end{aligned}
\qquad (2.2.25)
$$

In this case, the integrand for the first term (in parentheses) is an even function, so this term will not cancel out. But because the integrand for the second or time-retarded term is an odd function, the second term will cancel out (vanish). This is an important conclusion. To first order in $v/c_g$, the radial gravitational field in the rest frame of a massive central object is not time-retarded. That is why the classical Newtonian instantaneous action-at-a-distance inverse-square law (and not time-retarded) theory has been very successful for more than three centuries.

## 2.3 Review of the Empirical Formula for the Flyby Speed-Change

Let $\delta v_{emp}$ be the change in speed given by the empirical prediction formula found by J. D. Anderson et al.,[7] so that

$$\delta v_{emp} = \frac{2r_E \Omega_E}{c} v_\infty \left( \cos\left(\delta_{in}\right) - \cos\left(\delta_{out}\right)\right) , \qquad (2.3.1)$$

where $r_E$ is the Earth's spherical radius, $\Omega_E$ is the Earth's angular speed, $c$ is the vacuum speed of light, $v_\infty$ is the absolute magnitude of the asymptotic velocity vector at infinity, and $\delta_{in}$ and $\delta_{out}$ are the inbound and outbound asymptotic inertial declinations. This formula produces calculated flyby speed changes that are close to, but not exactly equal to, the observed speed changes for all of the reported Earth flybys.

The beauty of the empirical formula is its simplicity, but the problem is to identify the declinations, $\delta_{in}$ and $\delta_{out}$. J. D. Anderson et al. do not clearly define what they mean by "declination", but they do write, "The arcane difference between geocentric latitude and inertial declination is not statistically significant." (The appropriate word "arcane" means "mysterious" or "secret".) The traditional definition states that declinations north of the equatorial plane are *positive* and those south of the equatorial plane are *negative*.[12] For the NEAR flyby, the listed declination values are $\delta_{in}$=-20.76° and $\delta_{out}$=-71.96°. The calculated values for the geocentric latitude are $\lambda_{in}$=+20.82° and $\lambda_{out}$=-71.91°. It appears that J. D. Anderson et al. are using $\delta_{in}$=-$\lambda_{in}$ and $\delta_{out}$=$\lambda_{out}$. Of course, this curious sign reversal has no effect on the empirical formula, because $\cos(-\delta_{in})$=$\cos(\delta_{in})$.

Let's modify the empirical formula so that it is expressed in terms of the geocentric latitude, $\lambda$, and can be applied to a hypothetical flyby with a parabolic trajectory, for which $v_\infty$=0. Assume the inbound asymptotic geocentric latitude, $\lambda_{in}$, and outbound asymptotic geocentric latitude, $\lambda_{out}$, are known, and assume that $\lambda_{in}$ and $\lambda_{out}$ can be substituted for $\delta_{in}$ and $\delta_{out}$. Assume that $v_\infty$ can be replaced by $v_{in}$=$v(\lambda_{in})$. Then the empirical formula becomes

$$\delta v_{memp} = \frac{2r_E\Omega_E}{c} v_{in} \left( \cos(\lambda_{in}) - \cos(\lambda_{out}) \right)$$

$$= \frac{2r_E\Omega_E}{c} v_{in} \int_\lambda \sin(\lambda)d\lambda = \frac{2r_E\Omega_E}{c} v_{in} \int_t \sin(\lambda) \frac{d\lambda}{dt} dt$$

$$= \frac{2r_E\Omega_E}{c} v_{in} \left( \int_{t_{in}}^{t_p} \sin(\lambda(t)) \frac{v_\lambda}{r_\lambda} dt + \int_{t_p}^{t_{out}} \sin(\lambda(t)) \frac{v_\lambda}{r_\lambda} dt \right)$$

$$= \delta v_{in} + \delta v_{out} \quad , \tag{2.3.2}$$

where $t_p$ is the time at perigee (usually $t_p$=0). Note that $d\lambda$ may be positive, negative, or zero, but dt is <u>always</u> positive. This formula identifies the spacecraft's geocentric latitude, $\lambda$, and the latitudinal component of the spacecraft's velocity, $v_\lambda$, as important variables for calculating the speed change during a flyby.

There are three cases for which the empirical formula predicts a zero value for the flyby speed change: (1) a flyby with perigee over the equator, for which $\lambda_{out}$=-$\lambda_{in}$; (2) a flyby with perigee over either pole, for which $\lambda_{out}$=$\lambda_{in}$; (3) a flyby in the equatorial plane, for which $\lambda$=0.

The empirical formula makes another interesting prediction. Consider two polar flyby cases. Case1: inbound asymptotic latitude, $\lambda_{in}$=0, outbound asymptotic latitude, $\lambda_{out}$=-45°, deflection angle, DA=45°. Case2: $\lambda_{in}$=0, $\lambda_{out}$=-45°, DA=3×45°=135°. Both cases have the same $\lambda_{in}$ and $\lambda_{out}$, but Case2 has a deflection angle that is three times greater than that for Case1. Although a DA of 135° may not be dynamically possible, the empirical formula appears to be cyclical in the deflection angle.

The time-retarded theory developed herein will be consistent with these predictions of the empirical formula only if the Earth's transverse gravitational field causes a change in the component of the spacecraft's velocity that is directed along increasing $\lambda$.[13]

## 2.4 Time-Retarded Transverse Gravitational Field and Speed Change for a Spacecraft Orbiting a Large Rotating Sphere

Consider a spacecraft that is in a Keplerian orbit, revolving around a large rotating sphere (a spherical approximation for the Earth). Numerical values for the Earth's parameters are listed in Appendix A.

Let $(X,Y,Z)$ be a nonrotating geocentric frame of reference with the $(X,Y)$ plane being the equatorial plane and the $Z$-axis being the axis of rotation. Let $\lambda_p$ be the spacecraft's geocentric latitude at perigee. Let $(x,y,z)$ be the frame in which the $(x,y)$ plane is the orbital plane. Let the x-axis coincide with the X-axis if $\lambda_p=0$. Let $r$ be the geocentric radial distance to the field point in the $(x,y)$ plane, let $\varepsilon$ be the eccentricity for the orbit, let $\alpha_{eq}$ be the directed angle from the z-axis of the $(x,y,z)$ frame to the Z-axis of the $(X,Y,Z)$ frame, and let $\theta$ be the parametric angle for the field point in the plane of the orbit. Let $\theta$ range from $\theta_{min}$ to $\theta_{max}$. In this case, $\theta_{min}=-\pi$ and $\theta_{max}=+\pi$.

Let $E$ be the mechanical energy, the kinetic plus scalar potential energy, of the spacecraft, let $v$ be the orbital speed, let $M_E$ be the mass of the central sphere, and let $m_{sc}$ be the mass of the spacecraft. A good first approximation for $E$ is given by conservation of energy,

$$\text{constant} \cong E = KE + PE = \frac{1}{2}\,m_{sc}v^2 - \frac{GM_E m_{sc}}{r} = \frac{1}{2}\,m_{sc}v_p^2 - \frac{GM_E m_{sc}}{r_p} \quad , \qquad (2.4.1)$$

where $r_p$ and $v_p$ are the geocentric radial distance to the field point and the orbital speed at perigee ($\theta=0$), respectively. Solving for $v$ gives

$$v^2 = v_p^2 - \frac{2GM_E}{r_p} + \frac{2GM_E}{r} \quad ,$$

$$v(\theta) = \left( v_p^2 - \frac{2GM_E}{r(\theta)} \left( \frac{r(\theta)}{r_p} - 1 \right) \right)^{\frac{1}{2}} \quad . \qquad (2.4.2)$$

The value for $v_p$ can be found by equating the centripetal acceleration to the gravitational acceleration at perigee, such that

$$\frac{v_p^2}{r_p} = \frac{GM_E}{r_p^2} \quad , \quad v_p = \left( \frac{GM_E}{r_p} \right)^{\frac{1}{2}} \quad , \qquad v(\theta) = \left( \frac{GM_E}{r_p} \left( \frac{2r_p}{r(\theta)} - 1 \right) \right)^{\frac{1}{2}} \quad . \qquad (2.4.3)$$

The formulas for the radial distance for an elliptical orbit and its derivative are

$$r(\theta) = \frac{r_p\left(1 + \varepsilon\right)}{1 + \varepsilon\cos\left(\theta\right)} \quad ,$$

$$\frac{dr}{d\theta} = \frac{r_p\left(1 + \varepsilon\right)\varepsilon\sin\left(\theta\right)}{\left(1 + \varepsilon\cos\left(\theta\right)\right)^2} = \frac{r^2}{r_p}\frac{\varepsilon}{1 + \varepsilon}\sin\left(\theta\right) \quad . \qquad (2.4.4)$$

If the spacecraft is in a circular orbit, $\varepsilon=0$, $r=r_p$, $dr/d\theta=0$, and $v=v_p$.

Let $v_\perp$ be the component of $\mathbf{v}$ that is perpendicular to $\mathbf{r}$. Let $\Omega_\theta$ be the orbital angular speed, $\Omega_\theta \equiv d\theta/dt$. The formula for $v_\perp$ is

$$v_\perp = r\,\frac{d\theta}{dt} = r\Omega_\theta \quad . \tag{2.4.5}$$

Let $\mathbf{J}_{sc}$ be the angular momentum for the spacecraft. The vector $\mathbf{J}_{sc}$ is directed along the z-axis of the (x,y,z) frame. Let $\Omega_p$ be the spacecraft's angular speed at perigee, and let $J_p$ be the magnitude of its angular momentum at perigee. By conservation of angular momentum, the nearly constant absolute magnitude of $\mathbf{J}_{sc}$ is

$$\text{constant} \cong J_{sc} = m_{sc}v_\perp r = m_{sc}r^2\,\frac{d\theta}{dt} = m_{sc}r^2\Omega_\theta = J_p = m_{sc}r_p v_p = m_{sc}r_p^2\Omega_p \quad . \tag{2.4.6}$$

To a good first approximation, the formula for the angular speed in the orbital plane is

$$\Omega_\theta \equiv \frac{d\theta}{dt} = \frac{r_p v_p}{r^2} = \frac{r_p^2}{r^2}\,\Omega_p \quad . \tag{2.4.7}$$

Let $E_\theta$ be the angular energy of the spacecraft, and let $I_{sc}$ be the moment of inertia for the spacecraft. The spacecraft's angular energy is proportional to the ratio $I_{sc}/r^4$.

$$E_\theta = \frac{1}{2}\,I_{sc}\Omega_\theta^2 = \left(\frac{1}{2}\,r_p^2 v_p^2\right)\frac{I_{sc}}{r^4} \quad .$$

Let $\Delta t$ be the time interval relative to perigee at $\theta=0$. The formulas for $\Delta t$ and the orbital period P are

$$\Delta t(\theta) = \int_0^\theta \frac{dt}{d\theta}\,d\theta = \int_0^\theta \frac{1}{\Omega_\theta}\,d\theta \quad , \qquad P = \Delta t(\theta_{max}) - \Delta t(\theta_{min}) \quad . \tag{2.4.8}$$

Let $\theta_p$ be the parametric angle at which the orbit intersects the equatorial plane. The X, Y, and Z components of the geocentric radial distance to the field point, their time derivatives, and the components of $\mathbf{v}$ that are perpendicular to $\mathbf{r}$, are

$$r_X = r\cos(\theta - \theta_p) \quad ,$$
$$v_X = \frac{dr_X}{dt} = \left(\frac{dr}{d\theta}\cos(\theta - \theta_p)\right)\frac{d\theta}{dt} - \left(r\sin(\theta - \theta_p)\right)\frac{d\theta}{dt} \quad ,$$
$$vX_\perp = -r\Omega_\theta\sin(\theta - \theta_p) \quad ,$$
$$r_Y = r\cos(\alpha_{eq})\sin(\theta - \theta_p) \quad ,$$
$$v_Y = \frac{dr_Y}{dt} = \left(\frac{dr}{d\theta}\cos(\alpha_{eq})\sin(\theta - \theta_p)\right)\frac{d\theta}{dt} + \left(r\cos(\alpha_{eq})\cos(\theta - \theta_p)\right)\frac{d\theta}{dt} \quad ,$$
$$vY_\perp = r\Omega_\theta\cos(\alpha_{eq})\cos(\theta - \theta_p) \quad ,$$

$$r_z = -r \sin(\alpha_{eq}) \sin(\theta - \theta_p) \quad ,$$

$$v_z = \frac{dr_z}{dt} = -\left(\frac{dr}{d\theta} \sin(\alpha_{eq}) \sin(\theta - \theta_p)\right)\frac{d\theta}{dt} - \left(r \sin(\alpha_{eq}) \cos(\theta - \theta_p)\right)\frac{d\theta}{dt} \quad ,$$

$$vZ_\perp = -r\Omega_0 \sin(\alpha_{eq}) \cos(\theta - \theta_p) \quad . \tag{2.4.9}$$

Perigee is at $\theta=0$, and the orbit intersects the equatorial plane at $\theta=\theta_p$.

The value for the angle $\theta_p$ depends on the latitude for perigee $\lambda_p$, which ranges from $-90°$ to $+90°$, and $\alpha_{eq}$, which ranges from $0°$ to $180°$. If $\alpha_{eq}=0°$ or $180°$, then $\theta_p=0°$. If $\alpha_{eq}=90°$, then $\theta_p\propto\lambda_p$. If $0<\alpha_{eq}<\pi$ and $\alpha_{eq}\neq\pi/2$ and $\sin(\lambda_p)\leq\sin(\alpha_{eq})$, the formula for $\theta_p$ in radians is

$$\theta_p = \sin^{-1}\left(\frac{\sin(\lambda_p)}{\sin(\alpha_{eq})}\right) \quad . \tag{2.4.10}$$

If $\sin(\lambda_p)>\sin(\alpha_{eq})$, the inverse sine function is shifted from the primary branch and the value for $\theta_p$ is greater than $90°$. Of the six flybys reported by J. D. Anderson et al., only the MESSENGER flyby has a value for $\theta_p$ that is greater than $90°$; $\alpha_{eq}=133.1°$, $\lambda_p=46.95°$, and $\theta_p=90.0467°$. The MESSENGER flyby provides a serious challenge to any theory that purports to explain the flyby anomaly.

Let $\phi$ be the azimuthal angle for the projection of the field point onto the $(X,Y)$ equatorial plane, and let $\Omega_e$ be the angular speed for $\phi$, then

$$\Omega_e \equiv \frac{d\phi}{dt} = \frac{d\phi}{d\theta}\frac{d\theta}{dt} = \Omega_0 \frac{d\phi}{d\theta} \quad , \qquad\qquad \frac{d\phi}{d\theta} = \frac{\Omega_e}{\Omega_0} \quad . \tag{2.4.11}$$

Let $r_\phi$ be the geocentric radial distance to the projection of the field point onto the $(X,Y)$ equatorial plane, and let $r_\lambda$ be the geocentric radial distance to the projection of the field point onto the $(X,Z)$ plane. Thus,

$$r_\phi(\theta) = \left(r_x^2 + r_y^2\right)^{\frac{1}{2}} = r(\theta)\left(\cos^2(\theta - \theta_p) + \cos^2(\alpha_{eq}) \sin^2(\theta - \theta_p)\right)^{\frac{1}{2}} \quad ,$$

$$r_\lambda(\theta) = \left(r_x^2 + r_z^2\right)^{\frac{1}{2}} = r(\theta)\left(\cos^2(\theta - \theta_p) + \sin^2(\alpha_{eq}) \sin^2(\theta - \theta_p)\right)^{\frac{1}{2}} \quad . \tag{2.4.12}$$

The formula for the geocentric latitude, $\lambda$, is given by,

$$\lambda(\theta) = \tan^{-1}\left(\frac{r_z}{r_\phi}\right) = \pm \tan^{-1}\left(\frac{\sin(\alpha_{eq}) \sin(\theta - \theta_p)}{\left(\cos^2(\theta - \theta_p) + \cos^2(\alpha_{eq}) \sin^2(\theta - \theta_p)\right)^{\frac{1}{2}}}\right) \quad . \tag{2.4.13}$$

Let $v\phi_\perp$ be the component of $\mathbf{v}_\phi$ that is perpendicular to $\mathbf{r}_\phi$, such that

$$v\phi_\perp = r_\phi \frac{d\phi}{dt} = r_\phi\Omega_e \quad . \tag{2.4.14}$$

The formula for $v\phi_\perp$ is also given by the perpendicular components of **v** from Eq (2.4.9):

$$v\phi_\perp = \left(vX_\perp^2 + vY_\perp^2\right)^{\frac{1}{2}} = \pm r\Omega_0 \left(\sin^2(\theta - \theta_p) + \cos^2(\alpha_{eq}) \cos^2(\theta - \theta_p)\right)^{\frac{1}{2}} \quad .$$

Therefore,

$$\Omega_e = \frac{v\phi_\perp}{r_\phi} = \pm\Omega_0 \left(\frac{\sin^2(\theta - \theta_p) + \cos^2(\alpha_{eq}) \cos^2(\theta - \theta_p)}{\cos^2(\theta - \theta_p) + \cos^2(\alpha_{eq}) \sin^2(\theta - \theta_p)}\right)^{\frac{1}{2}} \quad ,$$

$$= \pm\Omega_0 \left(\frac{\cos^2(\alpha_{eq}) + \tan^2(\theta - \theta_p)}{1 + \cos^2(\alpha_{eq}) \tan^2(\theta - \theta_p)}\right)^{\frac{1}{2}} \quad . \tag{2.4.15}$$

Sign application: + if $\Omega_e r_\phi$ has the same direction as $\Omega_E r_E$; - if $\Omega_e r_\phi$ has the opposite direction as $\Omega_E r_E$. (The equatorial speed, $v_E = r_E\Omega_E$, is directed towards the east.)

The formula for $d\phi/d\theta$ is

$$\frac{d\phi}{d\theta} = \frac{\Omega_e}{\Omega_0} = \pm\left(\frac{\cos^2(\alpha_{eq}) + \tan^2(\theta - \theta_p)}{1 + \cos^2(\alpha_{eq}) \tan^2(\theta - \theta_p)}\right)^{\frac{1}{2}} \quad , \tag{2.4.16}$$

and the formula for $\phi(\theta)$ becomes

$$\phi(\theta) = \pm\int_0^\theta \left(\frac{\cos^2(\alpha_{eq}) + \tan^2(\theta - \theta_p)}{1 + \cos^2(\alpha_{eq}) \tan^2(\theta - \theta_p)}\right)^{\frac{1}{2}} d\theta \quad . \tag{2.4.17}$$

Use + if $\alpha_{eq} < 90°$; - if $\alpha_{eq} > 90°$.

The components of the radial distance to the spacecraft can now be rewritten in terms of the orthogonal variables r, $\lambda$, and $\phi$:

$$r_X = r \cos(\lambda) \cos(\phi) \quad ,$$

$$v_X = \frac{dr_X}{dt} = vX_r + vX_\lambda + vX_\phi = \left(\frac{dr}{d\theta} \cos(\lambda) \cos(\phi)\right)\frac{d\theta}{dt} - \left(r \frac{d\lambda}{d\theta} \sin(\lambda) \cos(\phi)\right)\frac{d\theta}{dt}$$
$$- \left(r \frac{d\phi}{d\theta} \cos(\lambda) \sin(\phi)\right)\frac{d\theta}{dt} \quad ,$$

$$r_Y = r \cos(\lambda) \sin(\phi) \quad ,$$

$$v_Y = \frac{dr_Y}{dt} = vY_r + vY_\lambda + vY_\phi = \left(\frac{dr}{d\theta} \cos(\lambda) \sin(\phi)\right)\frac{d\theta}{dt} - \left(r \frac{d\lambda}{d\theta} \sin(\lambda) \sin(\phi)\right)\frac{d\theta}{dt}$$
$$+ \left(r \frac{d\phi}{d\theta} \cos(\lambda) \cos(\phi)\right)\frac{d\theta}{dt} \quad ,$$

$$r_Z = r \sin(\lambda) \quad ,$$

$$v_Z = \frac{dr_Z}{dt} = vZ_r + vZ_\lambda = \left(\frac{dr}{d\theta} \sin(\lambda)\right)\frac{d\theta}{dt} + \left(r \frac{d\lambda}{d\theta} \cos(\lambda)\right)\frac{d\theta}{dt} \quad . \tag{2.4.18}$$

A reasonably valid formula for the Earth's mass-density distribution is given in Appendix C:

$$\rho(r) = \text{if}\left(r < r_{ic}, \rho_{ic}, \text{if}\left(r < r_{oc}, \rho_{oc}(r), \text{if}\left(r < r_{man}, \rho_{man}(r), \rho_{cst}(r)\right)\right)\right) \quad, \tag{C.2}$$

where $r_{ic}$ is the radius for the inner core, $r_{oc}$ is the outer radius for the outer core, $r_{man}$ is that for the mantle, and $r_E$ is the Earth's spherical radius. The mass densities for the interior annular parts are $\rho_{ic}$, $\rho_{oc}$, etc. Divide the sphere into a sequence of elemental point-mass sources, dm, with coordinates $(r, \lambda', \phi', t')$. The formula for dm is

$$dm = \rho(r)r^2 \cos(\lambda')dr d\lambda' d\phi' \quad. \tag{2.4.19}$$

The X, Y, and Z components of the radial distance for the source-point, $r$, are

$$\begin{aligned}
r_X &= r \cos(\lambda') \cos(\phi') \quad, \\
r_Y &= r \cos(\lambda') \sin(\phi') \quad, \\
r_Z &= r \sin(\lambda') \quad.
\end{aligned} \tag{2.4.20}$$

The X, Y, and Z components of the retarded distance, $r'$, are

$$\begin{aligned}
r'_X &= r_X - r_X = r \cos(\lambda) \cos(\phi) - r \cos(\lambda') \cos(\phi') \quad, \\
r'_Y &= r_Y - r_Y = r \cos(\lambda) \sin(\phi) - r \cos(\lambda') \sin(\phi') \quad, \\
r'_Z &= r_Z - r_Z = r \sin(\lambda) - r \sin(\lambda') \quad.
\end{aligned} \tag{2.4.21}$$

The magnitude of $r'$ is

$$\begin{aligned}
r' &= \left(r'^2_x + r'^2_y + r'^2_z\right)^{\frac{1}{2}} \\
&= \left(r^2 + r^2 - 2rr\left(\cos(\lambda)\cos(\lambda')\cos(\phi - \phi') + \sin(\lambda)\sin(\lambda')\right)\right)^{\frac{1}{2}} \quad,
\end{aligned}$$

which can be rewritten as

$$r' = r\left(1 + x\right)^{\frac{1}{2}} \quad, \tag{2.4.22}$$

where $x$ is defined by

$$x \equiv \frac{r^2}{r^2} - 2\frac{r}{r}\left(\cos(\lambda)\cos(\lambda')\cos(\phi - \phi') + \sin(\lambda)\sin(\lambda')\right) \quad. \tag{2.4.23}$$

The retarded-time derivative of $r'$ to first order in $d\phi/dt$ is

$$\frac{dr'}{dt'} = \frac{\partial r'}{\partial \phi}\frac{d\phi}{dt} + \frac{\partial r'}{\partial \phi'}\frac{d\phi'}{dt'} = \Omega_e \frac{\partial r'}{\partial \phi} + \Omega_E \frac{\partial r'}{\partial \phi'} \quad.$$

The formula for $dr'/dt'$ reduces to

$$\frac{dr'}{dt'} = \frac{rr}{r'}\left(\Omega_e - \Omega_E\right)\cos(\lambda)\cos(\lambda')\sin(\phi - \phi') \quad. \tag{2.4.24}$$

The transverse gravitational field is directed towards the east or west, which means that it cannot directly change the components of the field-point's velocity that are perpendicular to a unit vector directed towards the east. Therefore, to first order in $v/c_g$, the derivatives $(\partial r'/\partial r)(dr/dt)$ and $(\partial r'/\partial \lambda)(d\lambda/dt)$ cannot contribute to $g_e$.

The Jacobian-1 reduces to,

$$\frac{1}{c_g}\frac{dr'}{dt'} = \frac{r_E \Omega_E}{c_g}\frac{r}{r'}\frac{r}{r_E}\frac{\Omega_e - \Omega_E}{\Omega_E}\cos(\lambda)\cos(\lambda')\sin(\phi - \phi') \quad . \tag{2.4.25}$$

The transverse component of $r'$ is given by the vector cross product, $\mathbf{r} \times \mathbf{r'}/rr'$, such that

$$\mathbf{TC} = \frac{\mathbf{r} \times \mathbf{r'}}{rr'} = \frac{\mathbf{r} \times (\mathbf{r} - \mathbf{r})}{rr'} = \frac{\mathbf{r} \times \mathbf{r} - \mathbf{r} \times \mathbf{r}}{rr'} = \frac{0 - \mathbf{r} \times \mathbf{r}}{rr'} = \frac{\mathbf{r} \times \mathbf{r}}{rr'} \quad .$$

Written in component form,

$$\mathbf{TC} = \frac{1}{rr'}\left(\left(r_y r_z - r_z r_z\right)_x , \left(r_z r_x - r_x r_z\right)_y , \left(r_x r_y - r_y r_y\right)_z\right) \quad . \tag{2.4.26}$$

The Z-component of $\mathbf{TC}$ is

$$TC_z = \frac{1}{rr'}\left(r_x r_y - r_x r_y\right) = \frac{r}{r'}\cos(\lambda)\cos(\lambda')\sin(\phi - \phi') \quad . \tag{2.4.27}$$

Let $d^3g_e$ be the differential form for the time-retarded transverse gravitational field. The starting formula for $d^3g_e$ is

$$d^3g_e = \left(\frac{1}{r'^2} \text{ gravity law}\right)\left(TC_z\right)\left(\text{Jacobian-1}\right) \quad .$$

Substituting Eqs. (2.4.19), (2.4.25), and (2.4.27) into the starting formula then gives

$$d^3g_e = \left(-G\frac{\rho(r)r^2\cos(\lambda')dr d\lambda' d\phi'}{r'^2}\right)\left(\frac{r}{r'}\cos(\lambda)\cos(\lambda')\sin(\phi - \phi')\right)$$

$$\times \left(\frac{r_E \Omega_E}{c_g}\frac{r}{r'}\frac{r}{r_E}\frac{\Omega_e - \Omega_E}{\Omega_E}\cos(\lambda)\cos(\lambda')\sin(\phi - \phi')\right) \quad .$$

By rearrangement, $d^3g_e$ can be rewritten as the product of a coefficient, $A_g$, the ratio of the angular speeds, and the integrand, IG, which is

$$d^3g_e = -A_g\left(\frac{\Omega_e - \Omega_E}{\Omega_E}\right)IGd\phi'd\lambda'\frac{dr}{r_E} \quad , \tag{2.4.28}$$

where the definitions for the equatorial surface speed, $v_E$, and for the

coefficient $A_g$, are

$$A_g \equiv \left(\frac{v_E}{c_g}\right)\left(G\bar{\rho}_E r_E\right) \quad , \qquad v_E \equiv r_E\Omega_E \quad , \qquad (2.4.29)$$

and the integrand for the triple integration is defined by

$$IG \equiv \cos^2(\lambda)\left(\frac{\rho(r)}{\bar{\rho}_E}\frac{r^4}{r_E^4}\right)\left(\cos^3(\lambda')\right)\left(\frac{r_E}{r}\right)^3\left(\frac{\sin^2(\phi - \phi')}{(1 + x)^2}\right) \quad . \qquad (2.4.30)$$

Let $Ig\phi'$ be the integral over $\phi'$, let $Ig\lambda'$ be the integral over $\lambda'$, let $Igr$ be the integral over $r$, and let $Ig$ be the product $\cos^2(\lambda)\times Igr$, defined as follows.

$$Ig\phi' \equiv \left(\frac{r_E}{r}\right)^3 \int_{-\pi}^{\pi} \frac{\sin^2(\phi - \phi')}{(1 + x)^2}\,d\phi' \quad ,$$

$$Ig\lambda' \equiv \int_{-\pi/2}^{\pi/2} Ig\phi'\cos^3(\lambda')d\lambda' \quad ,$$

$$Igr \equiv \int_{0}^{r_E} Ig\lambda'\frac{\rho(r)}{\bar{\rho}_E}\frac{r^4}{r_E^4}\frac{dr}{r_E} \quad ,$$

$$Ig \equiv \cos^2(\lambda)Igr \quad . \qquad (2.4.31)$$

The triple integral, $Igr$ , can be solved by using numerical integration (I am using Mathcad14), but it takes a lot of computer time. To speed up the calculation, an algebraic expression for $Igr$ is needed.

By using numerical integration, it can be shown that $Igr$ is independent of $\lambda$ and $\phi$, which means that $Igr$ can be computed with $\lambda=0$ and $\phi=0$.

Assume that $Igr(r)$ can be approximated by a power series. Let $PSr(r)$ be a four-term power series, defined as follows.

$$PSr(r) \equiv \left(\frac{I_E}{\bar{\rho}_E r_E^5}\right)\left(\frac{r_E}{r}\right)^3\left(C_0 + C_2\left(\frac{r_E}{r}\right)^2 + C_4\left(\frac{r_E}{r}\right)^4 + C_6\left(\frac{r_E}{r}\right)^6\right) \quad . \qquad (2.4.32)$$

The coefficients can be adjusted to give an optimum fit to values for $Igr$ calculated by using numerical integration. By using a least-squares fitting routine, the following values for the coefficients were found to give an excellent fit of $PSr(r)$ to $Igr(r)$:

$$C_0 = 0.50889 \quad , \qquad C_2 = 0.13931 \quad ,$$
$$C_4 = 0.01013 \quad , \qquad C_6 = 0.14671 \quad . \qquad (2.4.33)$$

The quality of the fit using these coefficients is shown in Fig. 2. The maximum difference between $Igr$ and $PSr$ is less than $1\times10^{-5}$.

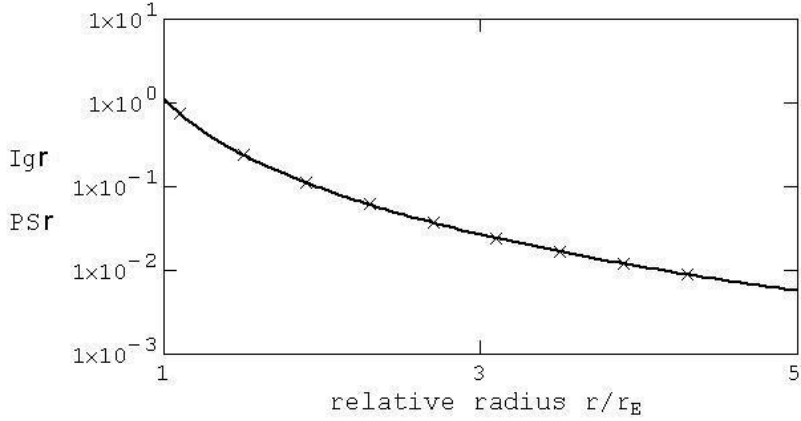

Figure 2. Semilog graph showing the excellent fit of $PSr(r)$ (solid curve) to calculated values for $Igr(r)$ (designated by ×). Notice that the power series can be extrapolated all the way down to the surface.

Let $PS(r)$ be defined as follows.

$$PS(r) \equiv \left(\frac{r_E}{r}\right)^3 \left(C_0 + C_2 \left(\frac{r_E}{r}\right)^2 + C_4 \left(\frac{r_E}{r}\right)^4 + C_6 \left(\frac{r_E}{r}\right)^6\right) \quad . \tag{2.4.34}$$

The solution for $g_e(\theta)$ can now be written as

$$g_e(\theta) = -A_e \left(\frac{\Omega_e(\theta) - \Omega_E}{\Omega_E}\right) \cos^2(\lambda(\theta)) PS(r(\theta)) \quad , \tag{2.4.35}$$

where the definition for the constant $A_e$ along with its numerical value (with $c_g=c$) is

$$A_e \equiv G \frac{I_E}{r_E^4} \frac{v_E}{c_g} = 5.0364 \times 10^{-6} \frac{m}{s^2} \quad . \tag{2.4.36}$$

Notice that $A_e$ is proportional to the ratio $I_E/r_E^4$. Let $E_E$ be the Earth's angular energy, for which the formula is

$$E_E = \frac{1}{2} I_E \Omega_E^2 \quad .$$

Recall that the spacecraft's angular energy, $E_{sc}$, is also proportional to the ratio $I_{sc}/r^4$.

Let $\mathbf{e}_r$, $\mathbf{e}_\lambda$, and $\mathbf{e}_\phi$ be orthogonal unit vectors for the spherical system $(r,\lambda,\phi)$; $\mathbf{e}_r$ is directed radially outward, $\mathbf{e}_\lambda$ is directed southward, and $\mathbf{e}_\phi$ is directed eastward. The polar angle or colatitude increases towards $+\mathbf{e}_\lambda$ (southward); the latitude $\lambda$ increases towards $-\mathbf{e}_\lambda$ (northward).

The vector velocity for the spacecraft in the $(r, \lambda, \phi)$ system becomes

$$\mathbf{v} = \mathbf{e}_r v_r + \mathbf{e}_\lambda v_\lambda + \mathbf{e}_\phi v_\phi \quad . \tag{2.4.37}$$

The transverse field $\mathbf{g}_e = -\mathbf{e}_\phi g_e$ is a solenoidal or vortex field.[14] The divergence of $\mathbf{g}_e$ is zero, which means it cannot be derived from the gradient of a scalar potential. But the curl of $\mathbf{g}_e$ is not zero, which means it can be derived from the curl of a vector potential. The modified empirical formula indicates that the speed change must be due to a change in the component of $\mathbf{v}$ that is directed along $\mathbf{e}_\lambda$. But the transverse field $\mathbf{g}_e$, being directed along $\mathbf{e}_\phi$, cannot directly change $\mathbf{v}_\lambda$. However, the time rate of change of $\mathbf{g}_e$ can by induction generate a vector field that is transverse to $\mathbf{g}_e$, i.e., directed along $\mathbf{e}_\lambda$. Induction fields are well-known in classical electromagnetic field theory (cf. Maxwell's Equations).[1]

Let $\mathbf{F}_\lambda$ be an induction field that is directed along $\mathbf{e}_\lambda$. Assume that the $\mathbf{e}_\phi$ component of the curl of $\mathbf{F}_\lambda$ equals $-k d\mathbf{g}_e/dt$, where k is a constant of proportionality. The formula for the curl operation in spherical coordinates can be found in J. D. Jackson's textbook.[1]

$$\nabla \times \mathbf{F}_\lambda = \mathbf{e}_\phi \frac{1}{r} \frac{\partial}{\partial r} \left( r F_\lambda \right) = -k \frac{d\mathbf{g}_e}{dt} = \mathbf{e}_\phi k \frac{dg_e}{dt} \quad .$$

Solving for $\partial(r F_\lambda)/\partial r$ and integrating both sides from t(0) to t($\theta$) gives

$$\int_{t(0)}^{t(\theta)} \frac{\partial}{\partial r} \left( r F_\lambda \right) dt = \int_0^\theta \frac{\partial}{\partial r} \left( r F_\lambda \right) \frac{dt}{d\theta} d\theta = \int_0^\theta \frac{d}{d\theta} \left( r F_\lambda \right) \frac{d\theta}{dr} \frac{dt}{d\theta} d\theta$$

$$= k \int_{t(0)}^{t(\theta)} r \frac{dg_e}{dt} dt = k \int_0^\theta r \frac{dg_e}{d\theta} d\theta \quad .$$

Therefore,

$$\int_0^\theta \left( \frac{d}{d\theta} \left( r F_\lambda \right) \frac{d\theta}{dr} \frac{dt}{d\theta} - kr \frac{dg_e}{d\theta} \right) d\theta = 0 \quad .$$

This equation is satisfied for all values of $\theta$, if and only if,

$$\frac{d}{d\theta} \left( r F_\lambda \right) = kr \frac{dr}{d\theta} \frac{d\theta}{dt} \frac{dg_e}{d\theta} \quad ,$$

so that,

$$F_\lambda(\theta) = \frac{k}{r(\theta)} \int_0^\theta r(\theta) \Omega_\theta(\theta) \frac{dr}{d\theta} \frac{dg_e}{d\theta} d\theta \quad . \tag{2.4.38}$$

This formula mathematically states the fundamental hypothesis for the time retarded theory. Units for $F_\lambda$ are $m/s^2$, which means that $F_\lambda$ has the same units as those for the radial field, $g_h$, and the transverse field, $g_e$. The factor k, which has units of $(m/s)^{-1}$, may be positive or negative. Let $v_k$ be the reciprocal of $|k|$, $v_k \equiv 1/|k|$. Regard k as an adjustable parameter and regard $v_k$ as a positive number that will be called the "induction speed". Let $v_E$ be the Earth's equatorial surface speed, $v_E \equiv r_E \Omega_E$. Regard $v_E$ as a convenient reference speed.

The formula for $F_\lambda$ can be rewritten as

$$F_\lambda(\theta) = \pm \frac{v_E}{v_k} \frac{r_E}{r(\theta)} \int_0^\theta \frac{r(\theta)}{r_E} \frac{\Omega_\theta(\theta)}{\Omega_E} \frac{1}{r_E} \frac{dr}{d\theta} \frac{dg_e}{d\theta} d\theta \quad . \tag{2.4.39}$$

The dot product $\mathbf{v} \cdot \mathbf{F}_\lambda$ changes the square of the magnitude of $\mathbf{v}_\lambda$, i.e., the spacecraft's kinetic energy. The dot product gives

$$\mathbf{v} \cdot \mathbf{F}_\lambda = \left( \mathbf{e}_r v_r + \mathbf{e}_\lambda v_\lambda + \mathbf{e}_\phi v_\phi \right) \cdot \mathbf{e}_\lambda F_\lambda = \pm v_\lambda F_\lambda = \pm r_\lambda \Omega_\theta F_\lambda \frac{d\lambda}{d\theta} \quad ,$$

because

$$v_\lambda = r_\lambda \frac{d\lambda}{dt} = r_\lambda \frac{d\lambda}{d\theta} \frac{d\theta}{dt} = r_\lambda \Omega_\theta \frac{d\lambda}{d\theta} \quad .$$

Choose $v_\lambda$ to be positive, then + if k is required to be positive and – if k is required to be negative.

This shows that $\mathbf{F}_\lambda$ changes the magnitude of $\mathbf{v}_\lambda$, as required by the modified empirical formula.

Now let's return momentarily to the modified empirical formula (Eq. 2.3.2).

$$\delta v_{memp} = \delta v_{in} + \delta v_{out} \quad ,$$

$$\delta v_{in} = \frac{2 r_E \Omega_E}{c} v_{in} \int_{t_{in}}^0 \sin(\lambda(t)) \frac{v_\lambda}{r_\lambda} dt = \frac{2 v_E}{c} v_{in} \int_{\theta_{min}}^0 \sin(\lambda(\theta)) \frac{d\lambda}{d\theta} d\theta \quad ,$$

$$\delta v_{out} = \frac{2 r_E \Omega_E}{c} v_{in} \int_0^{t_{out}} \sin(\lambda(t)) \frac{v_\lambda}{r_\lambda} dt = \frac{2 v_E}{c} v_{in} \int_0^{\theta_{max}} \sin(\lambda(\theta)) \frac{d\lambda}{d\theta} d\theta \quad , \tag{2.4.40}$$

where $v_{in} = v(\theta_{min})$. These formulas show that $\delta v_{in}$ and $\delta v_{out}$ depend on the parametric angles $\theta_{min}$ and $\theta_{max}$, respectively.

Let $\delta v/v_{in}$ be the relative change in the magnitude of $\mathbf{v}_\lambda$ due to $\mathbf{F}_\lambda$, where $v_{in}$ is the initial speed, such that

$$\left( 1 + \frac{\delta v}{v_{in}} \right)^2 \cong 1 + 2 \frac{\delta v}{v_{in}} = 1 + \frac{1}{v_{in}^2} \int_{t(0)}^{t(\theta)} F_\lambda v_\lambda dt = 1 + \frac{1}{v_{in}^2} \int_0^\theta r_\lambda F_\lambda \frac{d\lambda}{d\theta} d\theta \quad .$$

Therefore,

$$\delta v(\theta) = \frac{v_{in}}{2} \int_0^\theta \frac{r_\lambda(\theta) F_\lambda(\theta)}{v_{in}^2} \frac{d\lambda}{d\theta} d\theta \quad,$$

$$\delta v_{in} = \delta v(\theta_{min}) = \frac{v_{in}}{2} \int_0^{\theta_{min}} \frac{r_\lambda(\theta) F_\lambda(\theta)}{v_{in}^2} \frac{d\lambda}{d\theta} d\theta \quad,$$

$$\delta v_{out} = \delta v(\theta_{max}) = \frac{v_{in}}{2} \int_0^{\theta_{max}} \frac{r_\lambda(\theta) F_\lambda(\theta)}{v_{in}^2} \frac{d\lambda}{d\theta} d\theta \quad. \tag{2.4.41}$$

Let $\delta v_{trt}$ be the total speed change for the time retarded theory.

$$\delta v_{trt} = \delta v_{in} + \delta v_{out} \quad. \tag{2.4.42}$$

These formulas show that $\delta v_{trt}$ depends on $F_\lambda$, which depends on $dg_e/d\theta$, which in turn depends on the altitude, latitude, initial speed, and the initial value for the inclination angle $\alpha_{eq}$. For an equatorial orbit, $\cos(\alpha_{eq})=1$, $\lambda=0$, $d\lambda/d\theta=0$, and $\lambda_p=0$. In this case, $\delta v_{trt}=0$, i.e., there is no change in the magnitude of $\mathbf{v}_\lambda$, but there is a maximum change in angular momentum. At the other extreme, for a polar orbit, $\cos(\alpha_{eq})=0$, $\sin(\alpha_{eq})=1$, $\Omega_e=0$, $\theta_p=\lambda_p$, and $d\lambda/d\theta$ is at a maximum value. In this case, $\delta v_{trt}$ is at a maximum value, i.e., there is a maximum change in the magnitude of $\mathbf{v}_\lambda$. These results are consistent with the predictions of the modified empirical formula, Eq. (2.3.2).

The result of this heuristic exercise is a rational procedure for calculating the time-retarded speed change for spacecraft flybys of the Earth.

## 3. TRAJECTORY PARAMETERS FOR SPACECRAFT FLYBYS OF A LARGE SPINNING SPHERE

### 3.1 Trajectory Parameters in the Trajectory Plane

The given parameters for the NEAR spacecraft flyby will be used for the following calculation. Numerical values from the report of J. D. Anderson et al.[7] are listed in Appendix B. A schematic for the NEAR flyby is shown in Fig. 3. The same method used for the NEAR flyby will be applied to derive the time-retarded speed change for the remaining five flybys.

Let $r(\theta)$ be the geocentric radial distance to the spacecraft in the plane of the trajectory. The formulas for $r$ and its derivative are

$$r(\theta) = \frac{r_p(1+\varepsilon)}{1+\varepsilon\cos(\theta)} \quad, \qquad\qquad \frac{dr}{d\theta} = \frac{r(\theta)^2}{r_p} \frac{\varepsilon}{1+\varepsilon}\sin(\theta) \quad, \tag{3.1.1}$$

where $\theta$ is the parametric polar coordinate angle, $\varepsilon$ is the eccentricity for the hyperbolic trajectory, and $r_p$ is the radial distance to the spacecraft at perigee (at $\theta=0$).

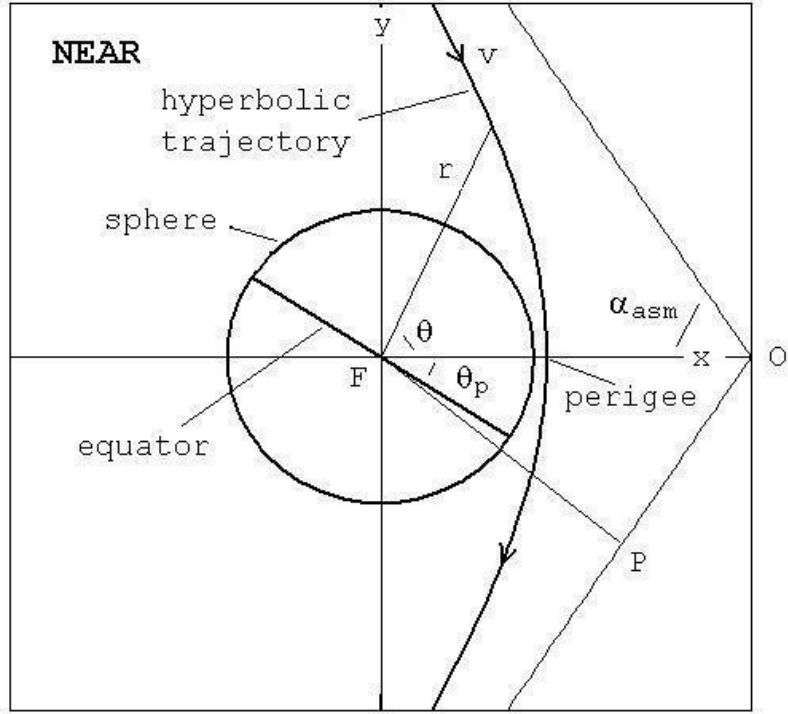

Figure 3. Hyperbolic trajectory for the NEAR spacecraft flyby in the (x,y) plane of the trajectory. The radial distance to the spacecraft is r at the parametric angle θ. The least distance $r_p$ is at perigee. The asymptotic angle $α_{asm}$ is defined by the deflection angle. The center of the sphere is at the focus F. The impact parameter is the distance FP. Another trajectory parameter is the distance OF. The spacecraft's speed is v.

The asymptotic angle $α_{asm}$ (Fig. 3) depends on the deflection angle,

$$α_{asm} = \frac{1}{2}\left(180° - DA\right) = \frac{1}{2}\left(180° - 66.9°\right) = 56.55° \ . \tag{3.1.2}$$

The radial distance at perigee, $r_p$, depends on the altitude at perigee,

$$r_p = r_E + h_p = r_E + 539\,km = 1.0846 r_E \ . \tag{3.1.3}$$

The impact parameter FP is given by conservation of angular momentum,

$$FPv_∞ = r_p v_p \ , \tag{3.1.4}$$

where $v_∞$ and $v_p$ are listed quantities and $r_p$ is given by Eq. (3.1.3). Numerical values for the NEAR flyby are

$$v_∞ = 6.851\,km/s \ , \quad v_p = 12.739\,km/s \ , \quad FP = r_p\frac{v_p}{v_∞} = 2.0167 r_E \ . \tag{3.1.5}$$

The ratio FP/OF=$\sin(\alpha_{asm})$. Therefore,

$$OF = \frac{FP}{\sin(\alpha_{asm})} = 2.4171 r_E \quad . \tag{3.1.6}$$

The parameter a is the distance OF-$r_p$,

$$a = OF - r_p = 1.3325 r_E \quad . \tag{3.1.7}$$

The parameter b depends on the asymptotic angle $\alpha_{asm}$,

$$b = a \tan(\alpha_{asm}) = 2.0170 r_E \quad . \tag{3.1.8}$$

The eccentricity $\varepsilon$ depends on a and b,

$$\varepsilon = \frac{\sqrt{a^2 + b^2}}{a} = 1.8142 \quad . \tag{3.1.9}$$

This gives the numerical value for $\varepsilon$ to be used in the formula for $r(\theta)$ (Eq. (3.1.1)).

The components of the radial distance for the field point in the new (x,y) trajectory plane, with perigee at $\theta=0$ and the equator at $\theta=\theta_p$, are

$$r_x(\theta) = r(\theta) \cos(\theta - \theta_p) \quad ,$$
$$r_y(\theta) = r(\theta) \sin(\theta - \theta_p) \quad . \tag{3.1.10}$$

Let (X,Y,Z) be a nonrotating geocentric frame of reference with the (X,Y) plane being the equatorial plane and the Z-axis being the axis of rotation. Then let the inclination of the plane of the trajectory to the equatorial plane be $\alpha_{eq}$. The value for $\theta_p$ depends on $\lambda_p$ and $\alpha_{eq}$. The formula for $\theta_p$ can be found by equating $r_y(0)$ to that component in the (X,Y) plane. Consequently,

$$r_p \sin(\theta_p) \sin(\alpha_{eq}) = r_p \sin(\lambda_p) \quad , \qquad \sin(\theta_p) = \frac{\sin(\lambda_p)}{\sin(\alpha_{eq})} \quad , \tag{3.1.11}$$

providing $\sin(\lambda_p) \le \sin(\alpha_{eq})$. For the NEAR spacecraft flyby, $\alpha_{eq}=108.0°$ and $\lambda_p=33.0°$, which gives

$$\theta_p = 34.9364° \quad . \tag{3.1.12}$$

A graph of the NEAR spacecraft flyby in the new (x,y) trajectory plane is shown in Fig. 4.

The X, Y, and Z components of the radial distance to the spacecraft

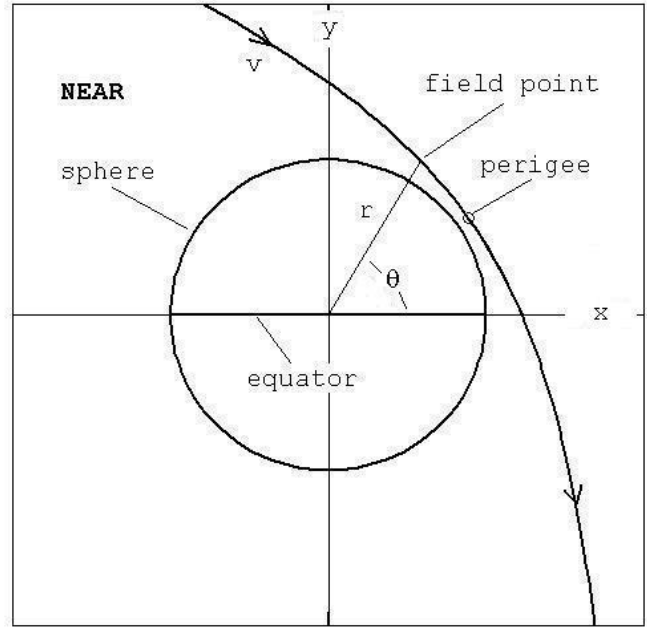

Figure 4. Trajectory for the NEAR spacecraft flyby in the new $(x, y)$ trajectory plane.

and the perpendicular components of the trajectory velocity are

$$r_X(\theta) = r(\theta) \cos(\theta - \theta_p) \; , \qquad\qquad vX_\perp(\theta) = -r(\theta)\Omega_\theta(\theta) \sin(\theta - \theta_p) \; ,$$
$$r_Y(\theta) = r(\theta) \cos(\alpha_{eq}) \sin(\theta - \theta_p) \; , \qquad vY_\perp(\theta) = r(\theta)\Omega_\theta(\theta) \cos(\alpha_{eq}) \cos(\theta - \theta_p) \; ,$$
$$r_Z(\theta) = -r(\theta) \sin(\alpha_{eq}) \sin(\theta - \theta_p) \; , \qquad vZ_\perp(\theta) = -r(\theta)\Omega_\theta(\theta) \sin(\alpha_{eq}) \cos(\theta - \theta_p) \; .$$

$$(3.1.13)$$

## 3.2 Linear Speed and Angular Speed of the Field Point During a Flyby

A good first approximation for the spacecraft's speed $v$ and the angular speed $\Omega_\theta$ can be found by using the classical conservation laws. By conservation of energy,

$$\frac{1}{2} v^2 - \frac{GM_E}{r} = \frac{1}{2} v_\infty^2 - \frac{GM_E}{r_\infty} = \frac{1}{2} v_\infty^2 \; ,$$

where $v_\infty$ is the listed speed at infinity, $r_\infty = \infty$, and $M_E$ is the Earth's mass. Solving for $v$ gives the formula for the trajectory speed,

$$v(\theta) = \left( v_\infty^2 + \frac{2GM_E}{r(\theta)} \right)^{\frac{1}{2}} \; . \tag{3.2.1}$$

Let $v_p$ be the speed at perigee. Compare the listed value for $v_p$ with the calculated value for $v(0)$, and the listed value for $v_\infty$ with the calculated value for $v(\theta_{min})$. (The value for $\theta_{min}$ will be found in Section 3.3.)

$$v_p = 12.739 \, km/s \; , \qquad\qquad v(0) = 12.742 \, km/s \; ,$$

$$v_\infty = 6.851\,\text{km/s} \quad , \qquad\qquad v_\text{in} = v(\theta_\text{min}) = 6.877\,\text{km/s} \quad .$$

The speed increase from $\theta_\text{min}$ to perigee for the NEAR flyby, $v_p/v_\text{in}=1.85$.

Let $\mathbf{J}_\text{sc}$ be the angular momentum vector of the spacecraft. By conservation of angular momentum, the magnitude of $\mathbf{J}_\text{sc}$ is

$$\text{constant} \equiv J_\text{sc} = m_\text{NEAR}v_\perp r = m_\text{NEAR}r^2\frac{dr}{dt} = m_\text{NEAR}r^2\Omega_\theta\,\frac{dr}{d\theta} = m_\text{NEAR}r_p v_p \quad .$$

Therefore,

$$\Omega_\theta(\theta) = \frac{r_p v_p}{r(\theta)^2} = \frac{r_p^{\,2}}{r(\theta)^2}\,\Omega_p \quad , \tag{3.2.2}$$

where $\Omega_p$ is the angular speed at perigee. Calculated values for $\Omega_\theta$ are

$$\Omega_p = \Omega_\theta(0) = \frac{v_p}{r_p} = 25.28\Omega_E \quad ,$$

$$\Omega_\text{in} = \Omega_\theta(\theta_\text{min}) = 2.456 \times 10^{-4}\Omega_E \quad ,$$

$$\Omega_p = 1.03 \times 10^5\Omega_\text{in} \quad . \tag{3.2.3}$$

The formula for $dt/d\theta$ is

$$\frac{dt}{d\theta} = \frac{1}{d\theta/dt} = \frac{1}{\Omega_\theta(\theta)} \quad . \tag{3.2.4}$$

## 3.3 Geocentric Latitude for the Field Point During a Flyby

The components of the radial distance for the field point in the nonrotating geocentric $(X,Y,Z)$ frame are given by Eq. (3.1.13). Let $r_\phi$ be the geocentric radial distance to the projection of the field point onto the $(X,Y)$ equatorial plane, and let $r_\lambda$ be the radial distance to the projection of the field point onto the $(X,Z)$ plane:

$$r_\phi(\theta) = \left(r_X^{\,2} + r_Y^{\,2}\right)^{\frac{1}{2}} = r(\theta)\left(\cos^2(\theta - \theta_p) + \cos^2(\alpha_\text{eq})\sin^2(\theta - \theta_p)\right)^{\frac{1}{2}} \quad ,$$

$$r_\lambda(\theta) = \left(r_X^{\,2} + r_Z^{\,2}\right)^{\frac{1}{2}} = r(\theta)\left(\cos^2(\theta - \theta_p) + \sin^2(\alpha_\text{eq})\sin^2(\theta - \theta_p)\right)^{\frac{1}{2}} \quad . \tag{3.3.1}$$

The geocentric latitude for the field point, $\lambda$, is

$$\lambda(\theta) = \tan^{-1}\left(\frac{r_z}{r_\phi}\right) = -\tan^{-1}\left(\frac{\sin(\alpha_\text{eq})\sin(\theta - \theta_p)}{\left(\cos^2(\theta - \theta_p) + \cos^2(\alpha_\text{eq})\sin^2(\theta - \theta_p)\right)^{\frac{1}{2}}}\right) \quad . \tag{3.3.2}$$

The listed values for the inbound and outbound declinations for the

NEAR flyby are

$$\delta_{in} = -20.76° \ , \qquad \delta_{out} = -71.96° \ . \tag{3.3.3}$$

Let $r_{in}$ be the radial distance at $\theta_{min}$, and let $r_{out}$ be the radial distance at $\theta_{max}$. (Values for $r_{in}$ and $r_{out}$ were not reported.)

If $r_{in}$ equals $348.5 r_E$, then

$$\theta_{min} = -\cos^{-1}\left(\frac{r_p(1 + \varepsilon)}{r_{in}\varepsilon} - \frac{1}{\varepsilon}\right) = -123.119° \ , \tag{3.3.4}$$

and

$$\lambda_{in} = \lambda(\theta_{min}) = +20.82° \ . \tag{3.3.5}$$

This value differs from $-\delta_{in}$ by 0.06°.

If $r_{out}$ equals $376.5 r_E$, then

$$\theta_{max} = +\cos^{-1}\left(\frac{r_p(1 + \varepsilon)}{r_{out}\varepsilon} - \frac{1}{\varepsilon}\right) = +123.144° \ , \tag{3.3.6}$$

and

$$\lambda_{out} = \lambda(\theta_{max}) = -71.91° \ . \tag{3.3.7}$$

This value differs from $\delta_{out}$ by 0.05°.

## 3.4 Time Interval for a Spacecraft Flyby

According to the report of J. D. Anderson et al.,[7] the data period for the NEAR flyby started 88.4 hours before perigee, and ended 95.6 hours after perigee. Calculated values for $\theta_{min}$ and $\theta_{max}$ are quite sensitive to the time interval for the data period.

Let $\Delta t(\theta)$ be the elapsed time for the data interval relative to perigee (at $\theta=0$). Therefore,

$$\Delta t(\theta) = \int_0^\theta \frac{dt}{d\theta}\, d\theta = \int_0^\theta \frac{1}{\Omega_0(\theta)}\, d\theta \ . \tag{3.4.1}$$

The values for $\Delta t(\theta_{min})$ and $\Delta t(\theta_{max})$, using the above values for $\theta_{min}$ and $\theta_{max}$ (Eqs. (3.3.4) and (3.3.6)), are

$$\Delta t\left(\theta_{min}\right) = -88.4 \text{ hours} \ ,$$
$$\Delta t\left(\theta_{max}\right) = +95.6 \text{ hours} \ , \tag{3.4.2}$$

which agree exactly with the reported data intervals.

If the values for $\theta_{min}$ and $\theta_{max}$ were

$$\theta_{min} = -123.1816° \ ,$$
$$\theta_{max} = +123.7159° \ , \tag{3.4.3}$$

the value for $\lambda_{in}$ would equal exactly $-\delta_{in}$ and the value for $\lambda_{out}$ would equal exactly $\delta_{out}$. But in this case the calculated values for the data intervals are way off;

$$\Delta t\left(\theta_{min}\right) = -109.4 \text{ hours } \ ,$$
$$\Delta t\left(\theta_{max}\right) \rightarrow \text{ does not converge } \ .$$

Obviously, calculated values for $\theta_{min}$ and $\theta_{max}$ can be more precisely determined from known values for the data intervals. Unfortunately, data intervals were not reported for flybys other than the NEAR flyby. Particularly useful would be the numerical values for the data intervals for the MESSENGER flyby.

## 3.5 Angular Speed for the Azimuthal Angle

Let $v\phi_\perp$ be the component of $\mathbf{v}_\phi$ that is perpendicular to $\mathbf{r}_\phi$,

$$v\phi_\perp = r_\phi \frac{d\phi}{dt} = r_\phi \Omega_e \ . \tag{3.5.1}$$

The formula for $v\phi_\perp$ is also given by the perpendicular components of $\mathbf{v}$ from Eq. (3.1.13):

$$v\phi_\perp = \left(vX_\perp^2 + vY_\perp^2\right)^{\frac{1}{2}} = -r\Omega_\theta \left(\sin^2(\theta - \theta_p) + \cos^2(\alpha_{eq})\cos^2(\theta - \theta_p)\right)^{\frac{1}{2}} \ .$$

Therefore,

$$\Omega_e = \frac{v\phi_\perp}{r_\phi} = -\Omega_\theta \left(\frac{\sin^2(\theta - \theta_p) + \cos^2(\alpha_{eq})\cos^2(\theta - \theta_p)}{\cos^2(\theta - \theta_p) + \cos^2(\alpha_{eq})\sin^2(\theta - \theta_p)}\right)^{\frac{1}{2}} \ ,$$
$$= -\Omega_\theta \left(\frac{\tan^2(\theta - \theta_p) + \cos^2(\alpha_{eq})}{1 + \cos^2(\alpha_{eq})\tan^2(\theta - \theta_p)}\right)^{\frac{1}{2}} \ . \tag{3.5.2}$$

Numerical values for the NEAR flyby are

$$v\phi_\perp(0) = -0.6264v_p \ , \qquad v\phi_\perp(\theta_{min}) = -2.7197 \times 10^{-3}v_{in} \ ,$$
$$\Omega_e(0) = -0.7467\Omega_\theta(0) \ , \qquad \Omega_e(\theta_{min}) = -0.5039\Omega_\theta(\theta_{min}) \ . \tag{3.5.3}$$

## 3.6 Speed Change for the NEAR Flyby Predicted by the Modified Empirical Formula

The spacecraft's initial speed, $v_{in}$, is

$$v_{in} = v(\theta_{min}) = 6.877 \text{ km/s } \ . \tag{3.6.1}$$

The modified version of the empirical formula (Eq. (2.3.2)) gives the following value.

$$\delta v_{memp} = \frac{2v_E}{c} v_{in} \left( \cos(\lambda_{out}) - \cos(\lambda_{in}) \right) = +13.33 \text{ mm/s} \quad . \tag{3.6.2}$$

This value is 1.1% less than the observed speed change.

$$\delta v_{obs} = +13.46 \text{ mm/s} \quad . \tag{3.6.3}$$

By using numerical differentiation and numerical integration (Eq. (2.4.39)), evaluation of the integral-differential equations for $\delta v_{in}$ and $\delta v_{out}$ gave the following values.

$$\delta v_{in} = \frac{2v_E}{c} v_{in} \int_{\theta_{min}}^{0} \sin(\lambda(\theta)) \frac{d\lambda}{d\theta} d\theta = +2.047 \frac{mm}{s} \quad ,$$

$$\delta v_{out} = \frac{2v_E}{c} v_{in} \int_{0}^{\theta_{max}} \sin(\lambda(\theta)) \frac{d\lambda}{d\theta} d\theta = +11.259 \frac{mm}{s} \quad , \tag{3.6.4}$$

$$\delta v_{in} + \delta v_{out} = +13.31 \text{ mm/s} \quad . \tag{3.6.5}$$

Both methods give essentially the same net speed change.

## 3.7 Components of the Radial Distance to the Field Point in the $(r,\lambda,\phi)$ System

The X, Y, and Z components of the radial distance to the spacecraft in terms of the orthogonal variables r, $\lambda$, and $\phi$ are

$$r_X = r \cos(\lambda) \cos(\phi) \quad ,$$
$$r_Y = r \cos(\lambda) \sin(\phi) \quad ,$$
$$r_Z = r \sin(\lambda) \quad . \tag{3.7.1}$$

The components of the trajectory velocity are

$$v_x = r\Omega_\theta \left( \frac{1}{r} \frac{dr}{d\theta} \cos(\lambda) \cos(\phi) - \frac{d\lambda}{d\theta} \sin(\lambda) \cos(\phi) - \frac{d\phi}{d\theta} \cos(\lambda) \sin(\phi) \right) \quad ,$$

$$v_y = r\Omega_\theta \left( \frac{1}{r} \frac{dr}{d\theta} \cos(\lambda) \sin(\phi) - \frac{d\lambda}{d\theta} \sin(\lambda) \sin(\phi) + \frac{d\phi}{d\theta} \cos(\lambda) \cos(\phi) \right) \quad ,$$

$$v_z = r\Omega_\theta \left( \frac{1}{r} \frac{dr}{d\theta} \sin(\lambda) - \frac{d\lambda}{d\theta} \cos(\lambda) \right) \quad . \tag{3.7.2}$$

Let $\mathbf{e}_r$, $\mathbf{e}_\lambda$, and $\mathbf{e}_\phi$ be orthogonal unit vectors for the spherical $(r,\lambda,\phi)$ system; $\mathbf{e}_r$ is directed radially outward, $\mathbf{e}_\lambda$ is directed southward, and $\mathbf{e}_\phi$ is directed eastward. The polar angle or colatitude increases towards $+\mathbf{e}_\lambda$ (southward); the latitude $\lambda$ increases towards $-\mathbf{e}_\lambda$ (northward).

The trajectory velocity in the $(r,\lambda,\phi)$ system is

$$\mathbf{v} = \mathbf{e}_r v_r + \mathbf{e}_\lambda v_\lambda + \mathbf{e}_\phi v_\phi \quad . \tag{3.7.3}$$

## 4. TIME-RETARDED TRANSVERSE GRAVITATIONAL FIELD AND SPEED CHANGE DURING A SPACECRAFT FLYBY OF A LARGE SPINNING SPHERE

### 4.1 Time-Retarded Transverse Gravitational Field for the NEAR Flyby

A good approximation for the Earth's mass-density distribution is given in Appendix C.

$$\rho(r) = \text{if}\left(r < r_{ic}, \rho_{ic}, \text{if}\left(r < r_{oc}, \rho_{oc}(r), \text{if}\left(r < r_{man}, \rho_{man}(r), \rho_{cst}(r)\right)\right)\right) \quad . \tag{C.2}$$

Divide the central sphere into a sequence of elemental point-mass sources, dm, with coordinates $(r,\lambda',\phi',t')$. The formula for dm is

$$dm = \rho(r)r^2 \cos(\lambda')drd\lambda'd\phi' \quad . \tag{4.1.1}$$

The X, Y, and Z components of the retarded distance, r', are

$$r'_X = r_X - \mathbf{r}_X = r\cos(\lambda)\cos(\phi) - \mathbf{r}\cos(\lambda')\cos(\phi') \quad ,$$
$$r'_Y = r_Y - \mathbf{r}_Y = r\cos(\lambda)\sin(\phi) - \mathbf{r}\cos(\lambda')\sin(\phi') \quad ,$$
$$r'_Z = r_Z - \mathbf{r}_Z = r\sin(\lambda) - \mathbf{r}\sin(\lambda') \quad . \tag{4.1.2}$$

The magnitude of r' is

$$r' = r\left(1 + x\right)^{\frac{1}{2}} \quad , \tag{4.1.3}$$

where $x$ is defined by,

$$x = \frac{\mathbf{r}^2}{r^2} - 2\frac{\mathbf{r}}{r}\left(\cos(\lambda)\cos(\lambda')\cos(\phi - \phi') + \sin(\lambda)\sin(\lambda')\right) \quad . \tag{4.1.4}$$

The retarded-time derivative of r' is

$$\frac{dr'}{dt'} = \frac{r\mathbf{r}}{r'}\left(\Omega_e - \Omega_E\right)\cos(\lambda)\cos(\lambda')\sin(\phi - \phi') \quad . \tag{4.1.5}$$

The Jacobian-1 is

$$\frac{1}{c_g}\frac{dr'}{dt'} = \frac{r_E \Omega_E}{c_g}\frac{r}{r'}\frac{\mathbf{r}}{r_E}\frac{\Omega_e - \Omega_E}{\Omega_E}\cos(\lambda)\cos(\lambda')\sin(\phi - \phi') \quad . \tag{4.1.6}$$

The Z-component of the relative cross product is

$$TC_Z = \frac{1}{rr'}\left(r_X r_Y - r_X \mathbf{r}_Y\right) = \frac{\mathbf{r}}{r'}\cos(\lambda)\cos(\lambda')\sin(\phi - \phi') \quad . \tag{4.1.7}$$

The differential $d^3g_e$ is

$$d^3g_e = -A_g \left( \frac{\Omega_e - \Omega_E}{\Omega_E} \right) IG d\phi' d\lambda' \frac{dr}{r_E} \quad , \tag{4.1.8}$$

where $A_g$ is defined by

$$A_g \equiv \left( G\bar{\rho}_E r_E \right) \left( \frac{v_E}{c_g} \right) \quad , \tag{4.1.9}$$

and the integrand for the triple integration is defined by

$$IG \equiv \cos^2(\lambda) \left( \frac{\rho(r)}{\bar{\rho}_E} \frac{r^4}{r_E^4} \frac{dr}{r_E} \right) \left( \cos^3(\lambda') d\lambda' \right) \left( \frac{r_E}{r} \right)^3 \left( \frac{\sin^2(\phi - \phi')}{(1 + x)^2} d\phi' \right) \quad . \tag{4.1.10}$$

Let $Ig\phi'$ be the integral over $\phi'$, let $Ig\lambda'$ be the integral over $\lambda'$, let $Igr$ be the integral over $r$, and let $Ig$ be the product $(-1)\cos^2(\lambda) \times Igr$, defined as follows.

$$Ig\phi' \equiv \left( \frac{r_E}{r} \right)^3 \int_{-\pi}^{\pi} \frac{\sin^2(\phi - \phi')}{(1 + x)^2} d\phi' \quad ,$$

$$Ig\lambda' \equiv \int_{-\pi/2}^{\pi/2} Ig\phi' \cos^3(\lambda') d\lambda' \quad ,$$

$$Igr \equiv \int_0^{r_E} Ig\lambda' \frac{\rho(r)}{\bar{\rho}_E} \frac{r^4}{r_E^5} dr \quad ,$$

$$Ig \equiv \cos^2(\lambda) Igr \quad . \tag{4.1.11}$$

By using numerical integration, it can be shown that $Igr$ is independent of $\lambda$ and $\phi$.

A good approximation for $Igr$ is the four-term power series $PSr$ given by Eq. (2.4.31):

$$Igr(r) = PSr(r) \equiv \left( \frac{I_E}{\bar{\rho}_E r_E^5} \right) \left( \frac{r_E}{r} \right)^3 \left( C_0 + C_2 \left( \frac{r_E}{r} \right)^2 + C_4 \left( \frac{r_E}{r} \right)^4 + C_6 \left( \frac{r_E}{r} \right)^6 \right) \quad , \tag{4.1.12}$$

where,

$$
\begin{array}{ll}
C_0 = 0.50889 \ , & C_2 = 0.13931 \ , \\
C_4 = 0.01013 \ , & C_6 = 0.14671 \ .
\end{array}
\tag{2.4.32}
$$

The formula for the transverse gravitational field becomes

$$g_e(\theta) = -A_e \left( \frac{\Omega_e(\theta) - \Omega_E}{\Omega_E} \right) \cos^2(\lambda(\theta)) PS(r(\theta)) \quad , \tag{4.1.13}$$

where

$$PS(r) \equiv \left(\frac{r_E}{r}\right)^3 \left(C_0 + C_2 \left(\frac{r_E}{r}\right)^2 + C_4 \left(\frac{r_E}{r}\right)^4 + C_6 \left(\frac{r_E}{r}\right)^6\right) \quad , \tag{4.1.14}$$

and the formula for $A_e$ along with its numerical value (with $c_g = c$) is

$$A_e \equiv G \frac{I_E}{r_E^4} \frac{v_E}{c_g} = 5.0364 \times 10^{-6} \frac{m}{s^2} \quad . \tag{4.1.15}$$

## 4.2 Time-Retarded Speed Change for the NEAR Spacecraft Flyby

The transverse gravitational field $\mathbf{g}_e$ is a vortex field that is directed towards $-\mathbf{e}_\phi g_e$, where $g_e$ is the signed magnitude of $\mathbf{g}_e$. The divergence for $\mathbf{g}_e$ is zero, which means $\mathbf{g}_e$ cannot be derived from the gradient of a scalar potential. But the curl is not zero, which means $\mathbf{g}_e$ can be derived from the curl of a vector potential. The modified empirical formula (Eq. (2.3.2)) indicates that to change the speed requires a change in the velocity component $\mathbf{v}_\lambda$. Although $\mathbf{g}_e$ cannot directly change $\mathbf{v}_\lambda$, the time-dependence of $\mathbf{g}_e$ can generate an induction field that is directed along $\mathbf{e}_\lambda$, and this induction field can change $\mathbf{v}_\lambda$.

Let $\mathbf{F}_\lambda$ be an induction field that is directed along $\mathbf{e}_\lambda$. Classical electromagnetic field theory suggests that the $\mathbf{e}_\phi$ component of the curl of $\mathbf{F}_\lambda$ equals $-kd\mathbf{g}_e/dt$, where $k$ is a constant of proportionality. The formula for the curl operator in spherical coordinates can be found in J. D. Jackson's textbook,[1] as

$$\nabla \times \mathbf{F}_\lambda = \mathbf{e}_\phi \frac{1}{r} \frac{\partial}{\partial r}\left(rF_\lambda\right) = \mathbf{e}_\phi k \frac{dg_e}{dt} \quad .$$

Solving for $F_\lambda$ gives (cf. Eq. (2.4.37))

$$F_\lambda(\theta) = \pm \frac{v_E}{v_k} \frac{r_E}{r(\theta)} \int_0^\theta \frac{r(\theta)}{r_E} \frac{\Omega_0(\theta)}{\Omega_E} \frac{1}{r_E} \frac{dr}{d\theta} \frac{dg_e}{d\theta} d\theta \quad ,$$
+ if k is positive, - if k is negative, $v_k \equiv 1/|k|$. (4.2.1)

A graph of $g_e(\theta)$ versus $\theta$, the derivative $dg_e/d\theta$ versus $\theta$, and the induction field $F_\lambda(\theta)$ versus $\theta$, with k positive, for the NEAR flyby, is shown in Fig. 5. The value used for $v_k$ to compute Fig. 5 was $4.130v_E$. The value used for $c_g$ was $1.060c$, which is the value found by Fomalont and Kopeikin for Jupiter.[4]

The dot product $\mathbf{v} \cdot \mathbf{F}_\lambda$ changes the spacecraft's kinetic energy from what it would be if $dg_e/d\theta$ were zero (circular orbit). Let $\delta v_{in}$ be the anomalous speed change that accumulates during the inbound angular interval from $\theta_{min}$ to 0, and let $\delta v_{out}$ be the anomalous speed change that

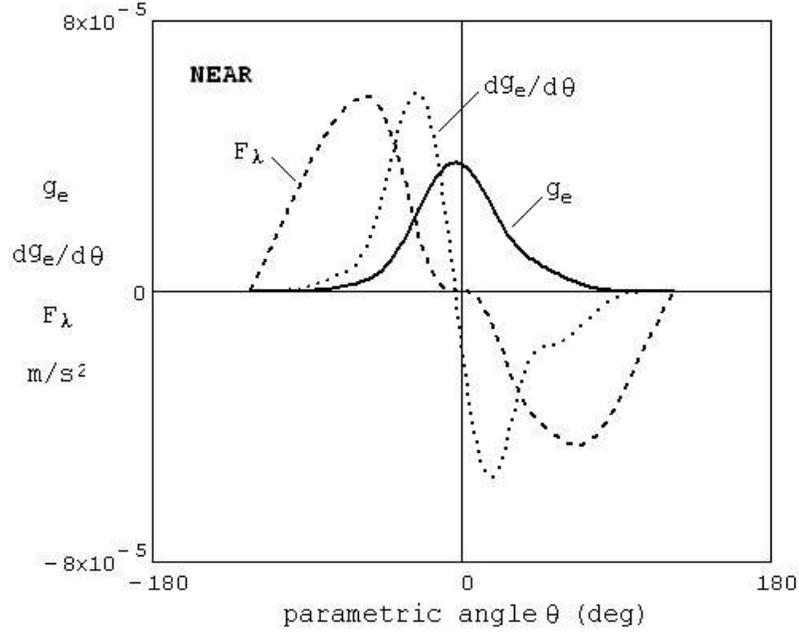

Figure 5. Time-retarded transverse gravitational field (solid curve), $g_e(\theta)$ versus $\theta$, the angular derivative of $g_e$ (dotted curve), $dg_e/d\theta$ versus $\theta$, and the induction field (dashed curve), $F_\lambda(\theta)$ versus $\theta$, for the NEAR flyby, using these values: $v_k=4.130v_E$, $c_g=1.060c$, and with k positive.

accumulates during the outbound angular interval from 0 to $\theta_{max}$ (cf. Eq. (2.4.40)):

$$\delta v_{in} = \frac{v_{in}}{2} \int_0^{\theta_{in}} \frac{r_\lambda F_\lambda}{v_{in}^2} \frac{d\lambda}{d\theta} d\theta \ ,$$

$$\delta v_{out} = \frac{v_{in}}{2} \int_0^{\theta_{max}} \frac{r_\lambda F_\lambda}{v_{in}^2} \frac{d\lambda}{d\theta} d\theta \ . \tag{4.2.2}$$

If $v_k/v_E=4.130\pm0.003$, and $c_g/c=1.060\pm0.001$, Eq. (4.2.2) gives

$$\delta v_{in} = -36.8988 \, \text{mm/s} \ ,$$

$$\delta v_{out} = +50.3589 \, \text{mm/s} \ . \tag{4.2.3}$$

The sum gives

$$\delta v_{trt} = \delta v_{in} + \delta v_{out} = +13.46 \pm 0.01 \, \text{mm/s} \ , \tag{4.2.4}$$

which agrees exactly with the observed speed change.

If $c_g$ is set equal to 1.000c, the value for $v_k/v_E$ that makes $\delta v_{trt}$ equal the observed speed change is 4.378±0.003. This value for $v_k$ is 6% greater than the value for $v_k$ with $c_g=1.060c$.

These results show that $c_g/c=1.060\pm0.001$ if the induction speed $v_k/v_E=4.130\pm0.003$. If the value for $v_k$ had been known with good precision, the NEAR flyby would put very stringent limits on the
possible range of values for $c_g$.

## 4.3 Anomalous Speed Change for the MESSENGER Flyby

For the MESSENGER flyby, the parametric angle for perigee, $\theta_p$, is very
close to +90°, and $\lambda_{out}$ is very close to $\lambda_{in}$. This means that the
anomalous speed change should be very small, and it is, only 0.02
mm/s.

A schematic of the trajectory for the MESSENGER spacecraft flyby, Fig.
6, shows a high degree of symmetry about perigee. Numerical values for
some listed parameters are

$$-\delta_{in} = -31.44° \ , \qquad \delta_{out} = -31.92° \ ,$$
$$\lambda_p = +46.95° \ , \qquad \alpha_{eq} = 133.1° \ ,$$
$$v_p = 10.389 \, \text{km/s} \qquad v_\infty = 4.056 \, \text{km/s} \ . \tag{4.3.1}$$

The reported speed change is

$$\delta v_{obs} = +0.02 \pm 0.01 \, \text{mm/s} \ . \tag{4.3.2}$$

Calculated values are

$$\varepsilon = 1.3596 \ , \qquad \theta_p = 90.0467° \ . \tag{4.3.3}$$

The time retarded speed change depends sensitively on the values used

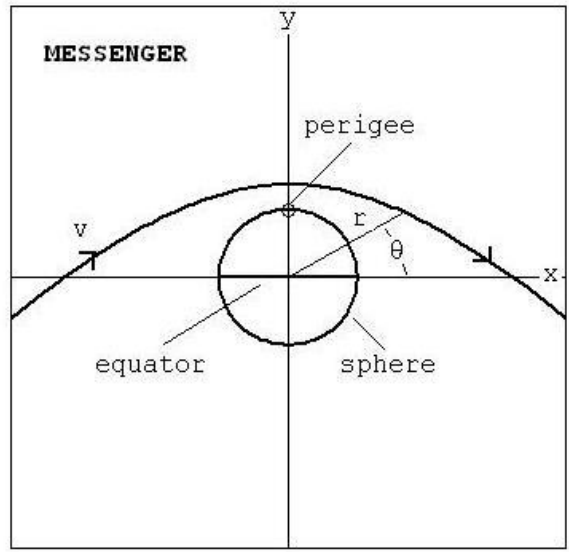

Figure 6. Trajectory for the MESSENGER spacecraft flyby in the (x,y) trajectory plane.

for $\theta_{min}$ and $\theta_{max}$. Assume that $\lambda(\theta_{min})=-\delta_{in}$. Then,

$$\theta_{min} = -135.4400° \ ,$$
$$\lambda(\theta_{min}) = -31.44° = -\delta_{in} \ ,$$
$$\Delta t(\theta_{min}) = -43.1\,hours \ . \tag{4.3.4}$$

The value for the inbound data interval with this value for $\theta_{min}$ seems too short.

Let $\theta_{min}$ have the value that will be used to calculate $\delta v_{trt}$.

$$\theta_{min} = -136.0625° \ ,$$
$$\lambda(\theta_{min}) = -31.75° \ ,$$
$$\Delta t(\theta_{min}) = -62.0\,hours \ . \tag{4.3.5}$$

This value for the inbound data interval seems about right, but the magnitude for the difference between $\lambda(\theta_{min})$ and $-\delta_{in}$ is 0.31°. To justify this value for $\theta_{min}$, let's assume the uncertainty in $\delta_{in}$ is at least ±0.31°,.

Now assume that $\lambda(\theta_{max})=\delta_{out}$. Then,

$$\theta_{max} = +136.4436° \ ,$$
$$\lambda(\theta_{max}) = -31.92° = \delta_{out} \ ,$$
$$\Delta t(\theta_{max}) = +90.1\,hours \ . \tag{4.3.6}$$

This value for the outbound data interval seems to be about right, and this is the value for $\theta_{max}$ that will be used to calculate $\delta v_{trt}$.

Let $\lambda_{in}=\lambda(\theta_{min})$, $\lambda_{out}=\lambda(\theta_{max})$, and $v_{in}=v(\theta_{min})=4.155$ km/s. Then the modified empirical formula gives

$$\delta v_{memp} = \frac{2v_E}{c}\,v_{in}\left(\cos(\lambda_{in}) - \cos(\lambda_{out})\right) = +0.02\,mm/s \ , \tag{4.3.7}$$

which agrees exactly with the observed speed change.

By adjusting the value for $v_k$, the time retarded theory can also produce exact agreement with the observed speed change. If $c_g/c=1.0\pm0.5$, $v_k/v_E=26\pm13$, and with k positive, the numerical values for $\delta v_{in}$, $\delta v_{out}$, and $\delta v_{trt}$, are

$$\delta v_{in} = \frac{v_{in}}{2}\int_0^{\theta_{min}} \frac{F_r F_\lambda}{v_{in}^2}\frac{d\lambda}{d\theta}\,d\theta = -5.632\,mm/s \ ,$$
$$\delta v_{out} = \frac{v_{in}}{2}\int_0^{\theta_{max}} \frac{F_r F_\lambda}{v_{in}^2}\frac{d\lambda}{d\theta}\,d\theta = +5.652\,mm/s \ ,$$
$$\delta v_{trt} = \delta v_{in} + \delta v_{out} = +0.02 \pm 0.01\,mm/s \ . \tag{4.3.8}$$

This value agrees exactly with the observed speed change.

## 4.4 Speed Change from the Anisotropic Light Speed Theory

A remarkable new theory based on a universal light-speed anisotropy and 3-space flow, reported by R. T. Cahill,[9] rejects a long held fundamental postulate of relativity theory, the invariance of the vacuum speed of light in all inertial frames of reference. To get the full basis for this new theory, one needs to review Cahill's voluminous report. In brief, by comparing the results of one-way speed of light experiments with the observed flyby speed change values, Cahill found the direction and magnitude for 3-space flow in the solar system and its effect on the speed of a spacecraft during Earth flybys. Only the calculated speed change will be noted here. Let $\delta v_{alt}$ be the speed change resulting from the anisotropic light-speed theory. The reported value for $\delta v_{alt}$ for the NEAR flyby is

$$\delta v_{alt} = 13.45 \, \text{mm/s} \quad , \tag{4.4.1}$$

which agrees very well with the observed speed change.

## 5. COMPARISON OF THE OBSERVED SPEED-CHANGE WITH CALCULATED VALUES FROM THE TIME-RETARDED THEORY AND OTHER METHODS FOR CALCULATING THE SPEED CHANGE

The values for $\theta_{min}$ and $\theta_{max}$ used to calculate $\delta v_{trt}$ and corresponding values for $\Delta t_{in}$, $\Delta t_{out}$, $\lambda_{in}$, and $\lambda_{out}$, are listed in Table I. For the NEAR flyby, $\theta_{min}$ and $\theta_{max}$ were adjusted to make $\Delta t_{in}$ and $\Delta t_{out}$ agree with the reported data intervals of 88.4 and 95.6 hours. For the remaining flybys, $\theta_{min}$ and $\theta_{max}$ were adjusted to make $\Delta t_{in}$ and $\Delta t_{out}$ produce what seemed to be reasonably valid data intervals, providing the calculated values for $\lambda_{in}$ and $\lambda_{out}$ were close to $-\delta_{in}$ and $\delta_{out}$, respectively. Those who created the flyby data should compare these estimated data intervals with the actually observed data intervals.

Table I. Values for the parametric angles $\theta_{min}$ and $\theta_{max}$ used to calculate the time-retarded speed change $\delta v_{trt}$ for each of the six flybys reported by J. D. Anderson et al.[7] Also listed are the calculated inbound and outbound data time intervals, $\Delta t_{in}$ and $\Delta t_{out}$, the geocentric latitudes at $\theta_{min}$ and $\theta_{max}$, $\lambda_{in}$ and $\lambda_{out}$, the trajectory speed at $\theta_{min}$, $v_{in}$, and the sign (+pos/-neg) used for k.

| Flyby | NEAR | GLL-I | Rosetta | M'GER | Cassini | GLL-II |
|---|---|---|---|---|---|---|
| $\theta_{min}$ (deg) | -123.119 | -113.625 | -138.641 | -136.063 | -99.689 | -115.339 |
| $\theta_{max}$ (deg) | +123.144 | +113.625 | +138.641 | +136.444 | +99.689 | +115.339 |
| $\Delta t_{in}$ (hours) | -88.4 | -88.2 | -85.2 | -62.0 | -55.1 | -88.9 |
| $\Delta t_{out}$ (hours) | +95.6 | +88.2 | +85.2 | +90.1 | +55.1 | +88.9 |
| $\lambda_{in}$ (deg) | +20.82 | +12.76 | +2.56 | -31.75 | +12.88 | +33.9 |
| $\lambda_{out}$ (deg) | -71.91 | -34.20 | -34.26 | -31.92 | -5.09 | -4.76 |
| $v_{in}$ (km/s) | 6.877 | 8.965 | 3.943 | 4.154 | 16.018 | 8.892 |
| k | pos | pos | neg | pos | pos | neg |

Table II. Comparison of the observed speed change for the six flybys reported by J. D. Anderson et al.,[7] $\delta v_{obs}$, the calculated time-retarded speed change, $\delta v_{trt}$, the ratio gravity-speed/light-speed, $c_g/c$, the characteristic induction speed required for the induction field, $v_k/v_E$, the eccentricity for the trajectory, $\varepsilon$, the speed-change from the modified empirical formula, $\delta v_{memp}$, and the reported value for the speed-change from the anisotropic light-speed theory,[9] $\delta v_{alt}$.

| Flyby | NEAR | GLL-I | Rosetta | M'GER | Cassini | GLL-II |
|---|---|---|---|---|---|---|
| $\delta v_{obs}$ (mm/s) | +13.46 ±0.01 | +3.92 ±0.30 | +1.80 ±0.03 | +0.02 ±0.01 | -2 ±1 | -4.6 ±1 |
| $\delta v_{trt}$ (mm/s) | +13.46 ±0.01 | +3.92 ±0.30 | +1.80 ±0.03 | +0.02 ±0.01 | -2 ±1 | -4.6 ±1 |
| $c_g/c$ | 1.060 ±0.001 | 1.0 ±0.1 | 1.06 ±0.02 | 1.0 ±0.5 | 1.0 ±0.5 | 1.0 ±0.2 |
| $v_k/v_E$ | 4.130 ±0.003 | 16 ±2 | 7.74 ±0.13 | 26 ±13 | 23 ±12 | 17 ±4 |
| $\varepsilon$ | 1.8142 | 2.4731 | 1.3122 | 1.3596 | 5.8456 | 2.3186 |
| $\delta v_{memp}$ (mm/s) | +13.31 | +4.1 | +2.11 | +0.02 | -1.06 | -4.7 |
| $\delta v_{alt}$ (mm/s) | +13.45 | +4.07 | +0.86 | -4.56 | -0.76 | -5.26 |

Results for the six flybys using the parameter values of Table I are listed in Table II. This table lists the observed anomalous speed change, $\delta v_{obs}$, with the reported uncertainty, the calculated time-retarded speed change, $\delta v_{trt}$, with the corresponding uncertainty, the ratio gravity-speed/light-speed, $c_g/c$, with the corresponding uncertainty, the required relative induction speed, $v_k/v_E$, with the corresponding uncertainty, the calculated value for the eccentricity for the trajectory, $\varepsilon$, the speed change from the modified empirical formula, $\delta v_{memp}$, and the reported speed change for the anisotropic light-speed theory, $\delta v_{alt}$.[9] The values for $v_k$ were adjusted to give exact agreement for $\delta v_{trt}$ with $\delta v_{obs}$.

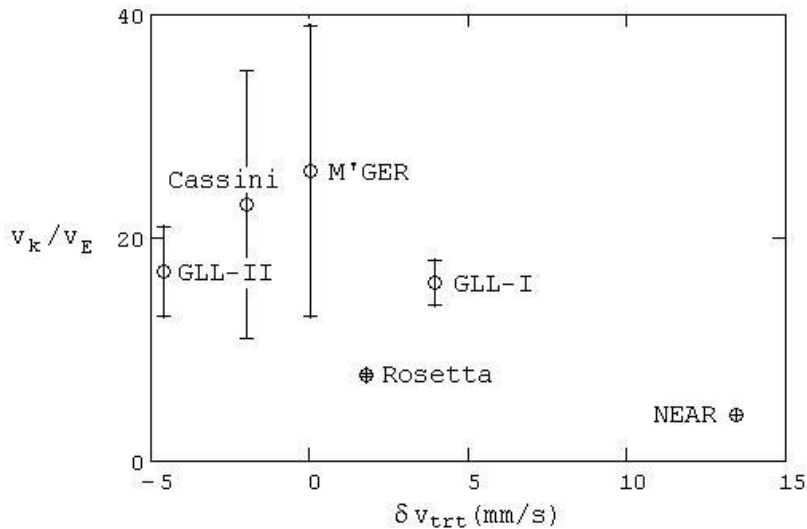

Figure 7. Required relative induction speed (designated by o), $v_k/v_E$, ± the associated uncertainty, versus the calculated time-retarded speed change, $\delta v_{trt}$, for each of the six flybys reported by J. D. Anderson et al.[7]

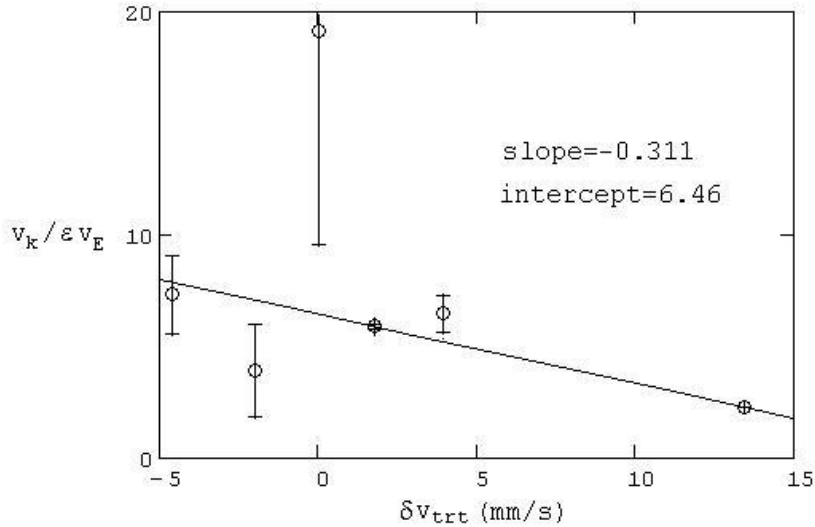

Figure 8. Linear regression line for $v_k/\varepsilon v_E$ versus $\delta v_{trt}$. This line fits exactly the NEAR and Rosetta flybys. An order of magnitude value for $v_k$ is given by $v_k \cong (6.46 - 0.311\delta v_{trt})\varepsilon v_E$, valid for $1 < \varepsilon < 6$ and $-5 < \delta v_{trt} < 15$ mm/s.

Notice in Table II that only the NEAR and Rosetta flybys have sufficient precision to identify a value for $c_g$ that differs significantly from c.

A graph of $v_k/v_E$ versus $\delta v_{trt}$, Fig. 7, shows an irregular dependence, but the general trend appears to be an increase in $v_k/v_E$ as $\delta v_{trt}$ decreases. A graph of $v_k/\varepsilon v_E$ versus $\delta v_{trt}$, Fig. 8, shows a nearly linear dependence. The formula for the regression line in Fig. 8 can be used to get an order-of-magnitude estimate for $v_k$.

$$v_k \cong \left(6.46 - 0.311\delta v_{trt}\right)\varepsilon v_E \approx 5v_E = 3\,\text{km/s} \quad . \tag{5.1.1}$$

The linear dependence of $v_k/\varepsilon v_E$ on $\delta v_{trt}$ may be caused by the Earth's oblateness.

## 6. CONCLUSIONS AND RECOMMENDATIONS

The concepts of time retardation and induction have been employed for more than a century in electromagnetic field theory. Gravitational theory, i.e., general relativity theory, reduces to classical electromagnetic field theory in the slow-speed weak-field approximation. It stands to reason that, in the slow-speed weak-field approximation, gravitational theory contains features that electromagnetic field theory contains, particularly time retardation and induction. The resulting neoclassical Newtonian field theory was used herein to derive a first order approximation for the speed change caused by the time dependence of the transverse gravitational field generated by a large spinning sphere (a simulation of the Earth) during spacecraft flybys. The value for the induction speed, $v_k$, was adjusted to make the value for the time-retarded speed change, $\delta v_{trt}$, agree exactly with the reported speed change, $\delta v_{obs}$. Results for the six flybys reported by J. D. Anderson et al.[7] are listed in Table II. This table shows conclusively that the time retarded theory can reproduce exact agreement with the facts of observation.

If the "true" value for the induction speed, $v_k$, had been known with good precision, the speed of the Earth's gravity field would have been firmly established, particularly from the data for the NEAR flyby. At this time, however, the time retarded theory produces results that are consistent with $c_g=1.000c$ or with $c_g=1.060c$.

The spherical model developed herein does not take into account nonspherical effects, such as the Earth's oblateness. A time retarded theory for a central rotating oblate spheroid could be developed, but that theory will be somewhat more complicated. Perhaps someday somebody will develop a suitable nonspherical theory.

Interesting is the similarity between the formula for $Ig\mathbf{r}$, Eq. (2.4.31), and the formula for $I_E$, Eq. (C.3).

$$Ig\mathbf{r} \equiv \int_0^{r_E} Ig\lambda' \frac{\rho(\mathbf{r})}{\bar{\rho}_E} \frac{r^4}{r_E^5} \, d\mathbf{r} \quad , \qquad I_E = \frac{8\pi}{3} \bar{\rho}_E r_E^5 \int_0^{r_E} \frac{\rho(\mathbf{r})}{\bar{\rho}_E} \frac{r^4}{r_E^5} \, d\mathbf{r} \quad .$$

Both integrals have the same dependence on $\rho(\mathbf{r})$ and $\mathbf{r}$. Table CI shows that more than 99% of $I_E$ is generated by the outer core and the mantle. If a ground-based instrument that can detect the Earth's transverse gravitational field were to be designed and built, it would confirm the time retarded theory and would be a useful tool for studying the structure of the mantle and outer core. Such an instrument has been described at arXiv:1009.3843 20Sep2010.[15]

## ACKNOWLEDGMENTS

I thank Patrick L. Ivers for reviewing the manuscript for this version (version 3) and suggesting improvements.

## APPENDIX A: USEFUL NUMERICAL VALUES FOR THE EARTH

Various numerical values are needed to evaluate the formulas for the Earth's transverse gravitational field. The following values were gleaned from several literature sources, mostly handbooks.

$G = 6.6732 \times 10^{-11} \ \dfrac{m^3}{kg \cdot s^2}$      Gravity constant

$c = 2.997925 \times 10^8 \ m/s$      Vacuum speed of light

$\Omega_E = 7.292115 \times 10^{-5} \ rad/s$      Earth's sidereal angular speed

$M_E = \left(5.9761 \pm 0.004\right) \times 10^{24} \ kg$      Earth's total mass

$r_e = 6\,378\,170 \pm 20$ m            Earth's equatorial radius

$I_{33} = \left(8.0413 \pm 0.0085\right) \times 10^{37}$ kg $\cdot$ m$^2$     Earth's polar moment of inertia

$I_{11} = \left(8.0150 \pm 0.0085\right) \times 10^{37}$ kg $\cdot$ m$^2$     Earth's equatorial moment of inertia

Computed values for other properties, where $r_p$ is the Earth's polar radius, are as follows.

$r_E = \left(r_e^2 r_p\right)^{\frac{1}{3}} = 6\,371\,034 \pm 21$ m      Earth's equivalent spherical radius

$v_E \equiv r_E \Omega_E = 464.58$ m/s            Earth's equatorial surface speed

$V_E = \left(1.08322 \pm 0.00001\right) \times 10^{21}$ m$^3$    Earth's volume

$\bar{\rho}_E = \left(5.517 \pm 0.004\right) \times 10^3$ kg/m$^3$     Earth's mean mass-density

$I_E = \dfrac{2}{3} I_{11} + \dfrac{1}{3} I_{33}$

$\quad = \left(8.0238 \pm 0.0085\right) \times 10^{37}$ kg $\cdot$ m$^2$    Earth's spherical moment of inertia

## APPENDIX B: PARAMETER VALUES FOR THE NEAR SPACECRAFT FLYBY

The following values are those listed in the report by J.D. Anderson et al. for the NEAR spacecraft flyby.[7] The uncertainty for the speed change is reported, but uncertainties for trajectory parameters were not reported.

$h_p = 539$ km                 altitude at perigee

$\lambda_p = 33.0$ deg             geocentric latitude at perigee

$v_p = 12.739$ km/s          spacecraft speed at perigee

$v_\infty = 6.851$ km/s           spacecraft speed at infinity

DA $= 66.9$ deg               deflection angle

$\alpha_{eq} = 108.0$ deg          inclination of trajectory to equator

$\delta_{in} = -20.76$ deg          inbound declination

$\delta_{out} = -71.96$ deg        outbound declination

$\delta v_{obs} = 13.46 \pm 0.01$ mm/s     observed change in spacecraft speed

## APPENDIX C: EARTH'S RADIAL MASS DENSITY DISTRIBUTION

Needed parameters and values for the Earth's radial mass density distribution can be found in various handbooks, such as the American Institute of Physics Handbook.[16] A cross section for the Earth's interior, depicted in Fig. C1, shows there are four major interior parts: inner core, outer core, mantle, and crust.

Let $r_{ic}$ be the radius for the inner core, $r_{oc}$ be the outer radius for the outer core, and $r_{man}$ be that for the mantle. Assume the outer radius for the crust is $r_E$. The following lists the outer radius (in meters) and the formula for the density (in kg/m³) for each component.

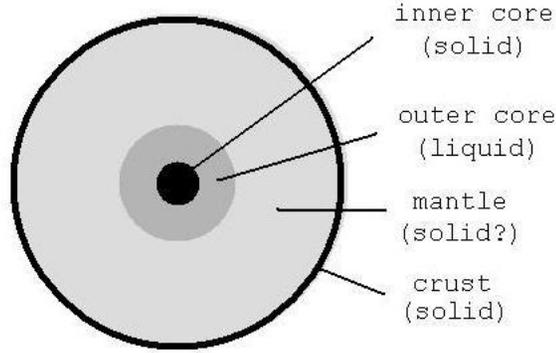

Figure C1. Cross section of the Earth's interior depicted as a solid spherical crustal shell containing a nearly solid mantle (highly viscous molten rock), a liquid (molten iron) outer core, and a solid (iron) inner core.

$$r_{ic} = 1230 \times 10^3 \ , \qquad \rho_{ic} = 13 \times 10^3 \ ,$$

$$r_{oc} = 3486 \times 10^3 \ ,$$

$$\rho_{oc}(r) = 12 \times 10^3 + 2.0 \times 10^3 \ \frac{(r_{ic} - r)}{(r_{oc} - r_{ic})} - 0.6 \times 10^3 \ \frac{(r_{ic} - r)^2}{(r_{oc} - r_{ic})^2} \ ,$$

$$r_{man} = 6321 \times 10^3 \ ,$$

$$\rho_{man} = 5.75 \times 10^3 + 0.4 \times 10^3 \ \frac{(r_{oc} - r)}{(r_{man} - r_{oc})} - 2.05 \times 10^3 \ \frac{(r_{oc} - r)^2}{(r_{man} - r_{oc})^2} \ ,$$

$$r_{cst} = r_E = 6378 \times 10^3 \ ,$$

$$\rho_{cst} = 3.3 \times 10^3 + 0.6 \times 10^3 \ \frac{(r_{man} - r)}{(r_{cst} - r_{man})} - 0.5 \times 10^3 \ \frac{(r_{man} - r)^2}{(r_{cst} - r_{man})^2} \ . \tag{C.1}$$

The formula for the radial density can be expressed by using a nested "if" function.

$$\rho(r) = \text{if}\left(r < r_{ic}, \rho_{ic}, \text{if}\left(r < r_{oc}, \rho_{oc}(r), \text{if}\left(r < r_{man}, \rho_{man}(r), \rho_{cst}(r)\right)\right)\right) \ . \tag{C.2}$$

A graph of $\rho(r)$ versus $r/r_E$ is shown in Fig. C2. This distribution gives the following values for the mass, $M_E$, and the spherical moment of inertia, $I_E$.

$$4\pi \bar{\rho}_E r_E^3 \int_0^{r_E} \frac{\rho(r)}{\bar{\rho}_E} \frac{r^2}{r_E^3} \ dr = 1.0072 M_E \ ,$$

$$\frac{8\pi}{3} \ \bar{\rho}_E r_E^5 \int_0^{r_E} \frac{\rho(r)}{\bar{\rho}_E} \frac{r^4}{r_E^5} \ dr = 1.0176 I_E \ . \tag{C.3}$$

These results suggest that the error in the formula for the Earth's mass density distribution, Eq. (C.2), probably is less than about 2%.

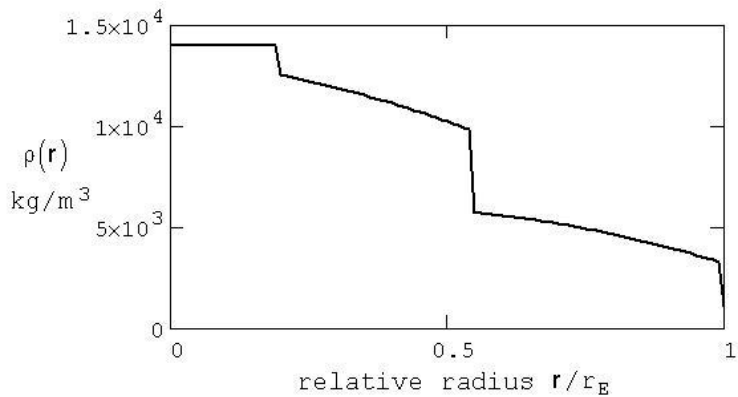

Figure C2. Radial mass density distribution for the Earth. The inner core, outer core, mantle, and crust are clearly evident in the step-like structure.

Interesting is the contribution of each interior part. Table CI lists the ratios of the Earth's interior parts to the total. Notice that the mass of the outer core plus the mass of the mantle is about 98% of the total mass, and the moment of inertia of the outer core plus the moment of inertia of the mantle is more than 99% of the total moment of inertia.

Table CI. Ratios of the Earth's interior parts to the total.

| $M_{ic}/M_E$ | $M_{oc}/M_E$ | $M_{man}/M_E$ | $M_{cst}/M_E$ | $I_{ic}/I_E$ | $I_{oc}/I_E$ | $I_{man}/I_E$ | $I_{cst}/I_E$ |
|---|---|---|---|---|---|---|---|
| 0.0170 | 0.2956 | 0.6826 | 0.0120 | 0.0008 | 0.1085 | 0.8844 | 0.0240 |